\documentclass[twocolumn]{aastex62}

\usepackage{amsmath}

\usepackage{longtable}

\usepackage{lineno}

\usepackage{float}

\newcommand{\tempo}{{\texttt{TEMPO}}}
\newcommand{\presto}{{\texttt{PRESTO}}}

\graphicspath{{./}{figures/}}

\received{Month, Day, Year}
\revised{Month, Day, Year}
\accepted{Month, Day, Year}

\submitjournal{ApJ}

\shorttitle{Timing and statistical analysis of PALFA sources}
\shortauthors{Doskoch et al.}

\begin{document}

\title{Timing and statistical analysis of single-pulse search pulsar discoveries from the PALFA survey}

\correspondingauthor{Graham\,M.\,Doskoch}
\email{gd00010@mix.wvu.edu}

\author[0000-0002-4219-6908]{Graham\,M.\,Doskoch}
\affiliation{Department of Physics and Astronomy, West Virginia University, Morgantown, WV 26506, USA}
\affiliation{Center for Gravitational Waves and Cosmology, West Virginia University, Chestnut Ridge Research Building, Morgantown, WV, 26505, USA}

\author[0000-0001-7697-7422]{Maura\,A.\,McLaughlin}
\affiliation{Department of Physics and Astronomy, West Virginia University, Morgantown, WV 26506, USA}
\affiliation{Center for Gravitational Waves and Cosmology, West Virginia University, Chestnut Ridge Research Building, Morgantown, WV, 26505, USA}

\author[0000-0003-4285-6256]{B.\,Allen}
\affiliation{Max-Planck-Institut für Gravitationsphysik (Albert-Einstein-Institut), D-30167 Hannover, Germany}
\affiliation{Department of Physics, University of Wisconsin - Milwaukee, Milwaukee WI 53211, USA}

\author[0000-0001-6341-7178]{A.\,Brazier}
\affiliation{Cornell Center for Astrophysics and Planetary Science, Ithaca, NY 14853, USA}

\author[0000-0002-1873-3718]{F.\,Camilo}
\affiliation{South African Radio Astronomy Observatory, Mowbray, 7705, South Africa}

\author{F.\,Cardoso}
\affiliation{Department of Physics and Astronomy, West Virginia University, Morgantown, WV 26506, USA}
\affiliation{Center for Gravitational Waves and Cosmology, West Virginia University, Chestnut Ridge Research Building, Morgantown, WV, 26505, USA}

\author[0000-0002-2878-1502]{S.\,Chatterjee}
\affiliation{Cornell Center for Astrophysics and Planetary Science, Ithaca, NY 14853, USA}

\author[0000-0002-4049-1882]{J. M.\,Cordes}
\affiliation{Cornell Center for Astrophysics and Planetary Science, Ithaca, NY 14853, USA}

\author[0000-0002-2578-0360]{F.\,Crawford}
\affiliation{Department of Physics and Astronomy, Franklin and Marshall College, Lancaster, PA 17604-3003, USA}

\author[0000-0003-1226-0793]{J. S.\,Deneva}
\affiliation{George Mason University, Fairfax, VA 22030, USA}

\author[0000-0002-2223-1235]{R. D.\,Ferdman}
\affiliation{School of Chemistry, University of East Anglia, Norwich Research Park, Norwich NR4 7TJ, UK}

\author[0000-0003-1307-9435]{P. C. C.\,Freire}
\affiliation{Max-Planck-Institut für Radioastronomie, Auf dem Hügel 69, Bonn 53121, Germany}

\author[0000-0003-2317-1446]{J. W. T.\,Hessels}
\affiliation{ASTRON, the Netherlands Institute for Radio Astronomy, Oude Hoogeveensedijk 4, 7991 PD Dwingeloo, The Netherlands}
\affiliation{Anton Pannekoek Institute for Astronomy, University of Amsterdam, Postbus 94249, 1090 GE Amsterdam, The Netherlands}

\author[0000-0001-9345-0307]{V.~M.~Kaspi}
\affiliation{Dept.~of Physics and McGill Space Institute, McGill Univ., Montreal, QC H3A 2T8, Canada}

\author[0000-0001-8503-6958]{J.\,van Leeuwen}
\affiliation{ASTRON, the Netherlands Institute for Radio Astronomy, Oude Hoogeveensedijk 4, 7991 PD Dwingeloo, The Netherlands}

\author[0000-0001-5229-7430]{R. S.\,Lynch}
\affiliation{Green Bank Observatory, P.O. Box 2, Green Bank, WV 24494, USA}

\author{A. G.\,Lyne}
\affiliation{Jodrell Bank Centre for Astrophysics, School of Physics and Astronomy, University of Manchester, Manchester, M13 9PL, UK}

\author[0000-0001-8845-1225]{B.~W.\,Meyers}
\affiliation{Australian SKA Regional Centre (AusSRC), Curtin University, Kent Street, Bentley, WA 6102, Australia}

\author[0000-0002-0430-6504]{E.\,Parent}
\affiliation{Université Grenoble Alpes, CNRS, IPAG, F-38000 Grenoble, France}  

\author[0000-0003-3367-1073]{C.\,Patel}
\affiliation{Dept.~of Physics and McGill Space Institute, McGill Univ., Montreal, QC H3A 2T8, Canada}
\affiliation{Dunlap Institute for Astronomy \& Astrophysics, University of Toronto, 50 St. George Street, Toronto, ON M5S 3H4, Canada}

\author[0000-0002-8509-5947]{B. B. P.\,Perera}
\affiliation{Arecibo Observatory, HC3 Box 53995, Arecibo, PR 00612, USA}

\author[0000-0001-5799-9714]{S. M.\,Ransom}
\affiliation{National Radio Astronomy Observatory, 520 Edgemont Rd., Charlottesville, VA 22903, USA}

\author[0000-0002-7374-7119]{P.\,Scholz}
\affiliation{Dunlap Institute for Astronomy \& Astrophysics, University of Toronto, 50 St. George Street, Toronto, ON M5S 3H4, Canada}

\author[0000-0001-9784-8670]{I. H.\,Stairs}
\affiliation{Dept. of Physics and Astronomy, University of British Columbia, 6224 Agricultural Road, Vancouver, BC V6T 1Z1, Canada}

\author[0000-0001-9242-7041]{B. W.\,Stappers}
\affiliation{Jodrell Bank Centre for Astrophysics, School of Physics and Astronomy, University of Manchester, Manchester, M13 9PL, UK}

\author[0000-0001-5105-4058]{W. W.\,Zhu}
\affiliation{CAS Key Laboratory of FAST, NAOC, Chinese Academy of Sciences, Beijing 100101, China}

\begin{abstract}
Almost two decades after their discovery, pulsars discoverable only through their single, dispersed radio pulses, known as rotating radio transients (RRATS), remain a poorly-understood class of objects. Compared to the overall pulsar population, few have timing solutions, limiting our ability to understand the mechanisms underlying their sporadic emission. Here, we present a single-pulse analysis of twelve sources from the PALFA survey, consisting of eleven objects initially identified as RRATs and one candidate fast radio burst. We present timing solutions for five of the sources with detections at a sufficient number of epochs and spin periods for two more. For all sources, we use a Bayesian framework to fit distributions of single-pulse energies, finding support for log-normal energy distributions and pulse-to-pulse wait times that are consistent with Poisson processes. Finally, we provide updates on the previously-published candidate fast radio burst J0613+18, with new indications that it is in fact extragalactic.
\end{abstract}

\section{Introduction} \label{sec:intro}

The past two decades of radio astronomy have seen the discovery of a plethora of fascinating objects detectable through their emission of single, dispersed radio pulses. Much attention has been given to fast radio bursts (FRBs; \citealt{Lorimer+2007}), millisecond-duration bursts of extragalactic origin. While the mechanisms behind FRBs are unknown \citep{Zhang2023}, magnetar flares have long been considered a possible culprit \citep{Popov+2010}, a model supported by the discovery of an extremely bright radio burst originating from a magnetar within the Milky Way \citep{Bochenek+2020}. On the opposite end of the transient duration scale lie long-period transients (LPTs; \citealt{Hurley-Walker+2022}), with emission periods ranging from tens of minutes to hours \citep{deRuiter+2024,Hurley-Walker2024}. As with FRBs, there are a variety of models of LPT emission, with M-dwarf/white dwarf binary systems an increasingly popular explanation thanks to optical observations \citep{deRuiter+2024,Hurley-Walker2024}.

At pulse widths comparable to FRBs and spin periods much shorter than LPTs lie rotating radio transients (RRATs; \citealt{McLaughlin+2006}). While there has been considerable debate about how to precisely define this class of sources, a popular definition is that RRATs are pulsars that were discovered only through their intermittent emission of sporadic single pulses, rather than through periodicity searches.\footnote{This is the definition used by, among others, \citealt{Keane+2011b}. One could also adopt the stronger definition that to be an RRAT, a source must be detectable \textit{only} through its single pulses and must not be detectable through periodicity searches.} Defining RRATs this way has the advantage of simplicity but the disadvantage of being dependent on the precise observing setup and search codes, rather than the intrinsic physical properties and behavior of the sources \citep{Keane+2011a}.

This definition could encompass multiple types of objects. One group is pulsars with faint ordinary emission that happen to exhibit extreme pulse-to-pulse modulation; another is neutron stars that are intrinsically sporadic emitters \citep{Burke-Spolaor+2010}. The former group could include more distant analogs to PSR~B0656+14, which emits strong single pulses but whose continuous emission is barely detectable \citep{Weltevrede+2006}. The latter group would require some mechanism to produce radio pulses only intermittently, or to attenuate or absorb  many radio pulses. Numerous proposals to  explain such intermittency have been put forth over the last two decades, including the fallback of supernova material \citep{Li2006}, radiation belts \citep{Luo+2007}, and asteroid belts or circumstellar material \citep{Cordes+2008}. A third group is extreme nullers, or objects that exhibit distinct ``on'' and ``off'' periods, with the on periods much rarer \citep{Wang+2007,Burke-Spolaor+2010,Burke-Spolaor2013}. It is quite plausible that the objects currently classified as RRATs are drawn from all three of these groups.

There are over 200 known RRATs. The RRATalog\footnote{\url{https://rratalog.github.io/rratalog/} } currently contains 336 sources \citep{2026arXiv260401203A}, while one review \citep{Abhishek+2022} identified 162 RRATs presented in the literature. More recently, the Galactic Plane Pulsar Snapshot survey (GPPS; \citealt{GPPS1})\footnote{\url{http://zmtt.bao.ac.cn/GPPS/}} on the Five Hundred Meter Aperture Spherical Telescope (FAST) reported over 70 new RRATs, or pulsars  discovered only through their single pulses \citep{GPPS2}. Interestingly, most of the sources presented in \citealt{GPPS2} fall into one of three categories: weak pulsars with sparse single pulses, pulsars exhibiting extreme nulling, and sources detected only through single pulses (termed ``proto-RRATs" by the authors). This supports the idea that RRATs are a complex, diverse group of objects.

RRATs remain mysterious because few have been well-studied to the same degree as canonical pulsars. Some have only a small number of detected pulses or have been detected at only a few epochs, which makes it difficult to obtain measurements of spin period P and period first derivative $\dot{P}$ through phase-coherent timing. These in turn enable the characterization of the RRAT and other pulsar populations, allowing for comparisons between the two (see e.g. \citealt{McLaughlin+2009}). Galactic surveys, combined with targeted follow-up observations and timing campaigns of candidates, provide the best means of improving our understanding of RRATs.

\begin{table*}
  \renewcommand\thetable{1}
  \centering
  \setlength{\tabcolsep}{1mm}
  \begin{tabular}{c c c c c c c c c}
  \hline
Source & Discovery date & Right ascension & Declination & DM & $P_{\mathrm{disc}}$ & $N_D$ & NE2025 distance & YMW16 distance \\
& (MJD) & (J2000) & (J2000) & (pc cm$^{-3}$) & (s) & & (kpc) & (kpc) \\
\hline\hline
J0529+25 & -- & 05:29:22 & +25:21 & 98.2 & -- & 1 & 2.2 & 1.8 \\
J0613+18 & 56974 & 06:13:00 & +18:47 & 400 & -- & 1 & $>14.1$ & $>25.0$ \\
J0623+15 & 57294 & 06:23:00 & +15:35 & 92.5 & -- & 1 & 2.2 & 1.7 \\
J0625+12 & 57733 & 06:25:19 & +12:52 & 102 & -- & 1 & 2.3 & 1.8 \\
J1843+05 & 58295 & 18:43:34 & +05:28 & 262.3 & 2.035 & 4 & 13.9 & 17.4 \\
J1859+07 & 57309 & 18:59:57 & +07:58 & 303 & -- & 1 & 9.4 & 11.4 \\
J1905+0413 & 57274 & 19:05:12 & +04:14 & 381 & 0.894 & 6 & 8.2 & 9.2 \\
J1906+0335 & 56398 & 19:06:54 & +03:36 & 212 & 1.296 & 4 & 6.0 & 6.1 \\
J1917+1142 & 55513 & 19:17:01 & +11:42 & 319 & 1.188 & 3 & 8.4 & 7.0 \\
J1924+10 & 58370 & 19:24:31 & +10:05 & 176.8 & 4.620 & 2 & 7.7 & 7.3 \\
J1929+1154 & 54907 & 19:29:13 & +11:54 & 80 & 3.218 & 4 & 4.1 & 2.6 \\
J2010+3147 & 54760 & 20:10:46 & +31:46 & 251 & 1.551 & 1 & 7.4 & 7.6 \\
\hline
\end{tabular}
{\footnotesize
  \caption{Discovery parameters of the twelve sources. $N_D$ is the number of pulses detected from each source in its discovery observation (or, in the case of J2010+3147, its second observation, which yielded enough pulses to estimate a period). The discovery observation for J0529+25 could not be retrieved. The right ascension and declination are the  coordinates of the center of the discovery beam. Estimated distances are based purely on the NE2025 and YMW16 electron density models and do not include halo contributions, which are expected to be important only for J0613+18. The rotation period of J1924+10 was estimated from its final observation instead of its discovery observation due to a lack of single pulses, but is included here for completeness.}}
\end{table*} \label{tab:discoveries}

The Pulsar Arecibo L-band Feed Array survey (PALFA; \citealt{Cordes+2006}) was carried out using the 305-meter William E. Gordon Telescope at the Arecibo Observatory from 2004 to 2020. As with many Galactic pulsar surveys, it targeted the Galactic plane ($|b|<5^{\circ}$), focusing in particular on the Galactic center and Galactic anticenter. Among its discoveries are over 200 pulsars (see e.g. \citealt{Parent+2022}), approximately 15 RRATs \citep{Deneva+2009,Patel+2018}, and three FRBs \citep{Spitler+2014,Patel+2018,ParentTNS}, including the first repeating FRB \citep{Spitler+2016}.\footnote{For an incomplete list of PALFA discoveries, see \url{https://palfa.nanograv.org/}. The bulk of the PALFA website is no longer hosted by the National Radio Astronomy Observatory, due to the closure of the Arecibo Observatory. Old versions of relevant web pages can be found on the Internet Archive at \url{https://web.archive.org/web/20211101135226/http://www2.naic.edu/alfa/pulsar/}.}

Here, we present an analysis of twelve sources discovered by the PALFA collaboration only through  single-pulse searches. Eleven of the sources were initially classified as RRATs, and one was identified as a candidate fast radio burst. Discovery observations of five were previously published in \citealt{Patel+2018}. In Section~\ref{sec:observations-and-sources}, we describe the discovery and follow-up observations and introduce the twelve sources. We describe our single-pulse search pipeline in Section~\ref{sec:search}. We present the results of our timing analysis in Section~\ref{sec:timing}. Sections~\ref{sec:morphology}, ~\ref{sec:pulse-energies} and ~\ref{sec:wait-times} describe various aspects of our single-pulse analysis. We discuss the implications of our results in Sections~\ref{sec:discussion} and conclude in Section~\ref{sec:conclusion}.

\section{Observations and sources} \label{sec:observations-and-sources}

\subsection{Survey observations} \label{subsec:survey-observations}

Using the 305-meter telescope at the Arecibo Observatory, the PALFA survey observed two portions of the Galactic plane: one spanning longitudes of $32^{\circ}\leq l\leq77^{\circ}$ (the ``inner" region) and one spanning longitudes of $168^{\circ}\leq l\leq214^{\circ}$ (the ``outer" region, including the Galactic anticenter). Survey observations were performed at L-band using the Arecibo L-band Feed Array (ALFA) receiver, producing seven beams with full-width at half-max of 3.6 arcminutes. The survey initially used the Wide-Band Arecibo Pulsar Processor spectrometers (WAPP; \citealt{Dowd+2000}) as a backend, but switched to the Mock spectrometers in 2009. Integration times were 268 seconds for the inner region and 180 seconds for the outer. These integration times are short relative to the mean time between pulses from many of the sources it discovered; as such, some of its single-pulse discoveries were initially only known from a handful of pulses. Further details on PALFA observing strategies can be found elsewhere in the literature \citep{Cordes+2006,Deneva+2009,Swiggum+2014,Lazarus+2015,Parent+2018,Patel+2018,Parent+2022}.


The WAPP backend spanned 100 MHz of bandwidth, centered at 1440 MHz, which was split into 256 frequency channels with channel bandwidth $\Delta f_{\mathrm{chan}}=390$ kHz. The data were taken with a sampling time of $65$ $\mu$s, then written to disk in PSRFITS format \citep{Hotan+2004}

The Mock spectrometers covered two overlapping subbands, each covering 172 MHz and split into 512 frequency channels ($\Delta f_{\mathrm{chan}}=336$ kHz), sampled at 65.476 $\mu$s. When combined, the subbands produced a single band of bandwidth $\Delta f=323$ MHz and 960 channels, centered at 1375 MHz. The data were then written to disk in PSRFITS format.\footnote{For detailed information on PALFA's use of the Mock spectrometers, see \citealt{Lazarus+2015}.}

Both periodicity and single-pulse searches were performed on each observation. Single-pulse candidates were identified using pipelines built on the \presto\footnote{\url{https://github.com/scottransom/presto}} package \citep{Ransom2011}. After the removal of radio frequency interference (RFI), the data were dedispersed over a range of dispersion measures (DMs), with a maximum of 10,000 pc cm$^{-3}$. Single-pulse candidates were then identified using a matched filter algorithm, and eventually inspected manually. Various upgrades and modifications were made to this process over the years, such as the precise dedispersion plans used. The first two versions of this pipeline are detailed in \citealt{Deneva+2009} and \citealt{Lazarus+2015}. A later pipeline \citep{Patel+2018} added clustering analysis and machine learning techniques to better identify astrophysical sources.

\subsection{Follow-up campaigns}

\subsubsection{Arecibo}

Once a candidate was identified as likely astrophysical, follow-up observations were performed at Arecibo. These observations were largely performed with the L-wide band (LWB) receiver and the new Puerto Rico Ultimate Pulsar Processing Instrument (PUPPI) spectrometer at 1380 MHz. PUPPI provided 800 MHz of bandwidth channelized into 2048 channels ($\Delta f_{\mathrm{chan}}=388$ kHz), of which roughly 600 MHz was typically useful after excision of interference. The data were sampled with a sampling time of 40.96 $\mu$s. In rare cases, a source was targeted by multiple pointings with the ALFA/Mock combination. The majority of these follow-up lasted 5, 10, 15 or 20 minutes, with some variation due to miscellaneous issues. Cable failures and the subsequent collapse of the telescope in late 2020 brought the PALFA survey to an abrupt end and interrupted follow-up campaigns. Further analysis of the data, however, is ongoing.

\subsubsection{CHIME}

Additional follow-up observations were taken using the CHIME telescope, with each observation lasting approximately 14 minutes, targeting in particular several of the more sporadic sources in the sample. Data were taken using the CHIME/Pulsar backend \citep{CHIME-Pulsar}, featuring 400 MHz of bandwidth centered at 600 MHz divided into 1024 frequency channels and sampled every 327.68 $\mu$s. The data were recorded in filterbank format.

\subsection{Sources} \label{subsec:sources}

\begin{figure}
    \centering
    \includegraphics[width=0.9\columnwidth]{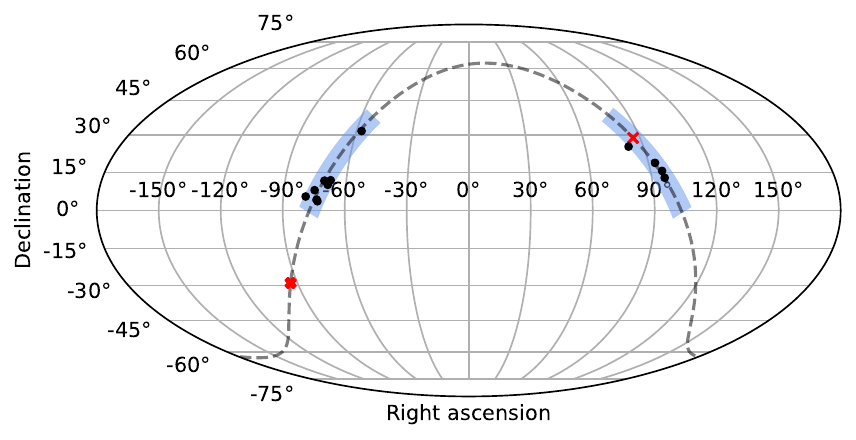}
    \caption{A skymap of the PALFA survey, featuring the regions surveyed (blue), the twelve sources discussed in this paper (black dots), the Galactic center (thick black cross) and Galactic anticenter (thin black cross). The ``inner'' region of the survey does not reach the Galactic center, which lies outside the declination range that was visible to the telescope.}
    \label{fig:skymap}
\end{figure}

Table~\ref{tab:discoveries} lists information about the twelve single-pulse sources discussed in this paper. J2010+3147 was discovered in an observation from 2008 using the WAPP backend, while the other eleven were discovered using the Mock spectrometers. As noted above, the discoveries of five of these sources (J0613+18, J0623+15, J0625+12, J1859+07, and J1905+0413) have already been published \citep{Patel+2018}. At the time of that paper's publication, follow-up campaigns had not yet concluded; as such, much could not yet be said about these objects.

J0613+18 was initially classified as a fast radio burst candidate \citep{Patel+2018} because of its high dispersion measure of $\sim400$ pc cm$^{-3}$, well in excess of the predicted maximum Galactic DM along that line of sight by either the NE2001 (188.3 pc cm$^{-3}$; \citealt{NE2001}) or YMW16 (297.1 pc cm$^{-3}$; \citealt{YMW16}) models.\footnote{The discrepancy is also present in the update to NE2001, NE2025 (159.1 pc cm$^{-3}$; \citealt{NE2025}).}  It was subsequently designated FRB 141113. A discussion of its discovery observation and a preliminary analysis can be found in \citealt{Patel+2018}. We continue this discussion in Section~\ref{subsec:J0613+18-nature}.

J1843+05 and J1924+10 were first detected in the summer of 2018. Like all PALFA discoveries, follow-up observations were conducted; however, the loss of the 305-meter telescope meant that follow-up campaigns targeting this pair of objects were cut short, with only 25 minutes of on-source data over three epochs taken for each.

For those discovery observations featuring three or more single pulses, we estimated rotation periods using the \presto\;utility \texttt{rrat\_period}, which computes a rotation period based on a set of arrival times by calculating the maximum possible divisor of the differences between times. These periods are also listed in Table~\ref{tab:discoveries}, and are largely quite typical of the RRAT population.

After their discoveries and preliminary announcement on the PALFA website, ten of these sources were redetected by the Galactic Plane Pulsar Snapshot (GPSS) survey with the FAST telescope. They were detected in either periodicity searches or single-pulse searches \citep{GPPS1,GPPS2,GPPS6,GPPS8} and were accordingly classified by GPPS as ordinary pulsars, RRATs, or ``weak pulsars with strong single pulses" \citep{GPPS2}. We reserve a comparison of the PALFA and GPPS results for Section~\ref{subsec:GPPS}, and then discuss the implications for rotating radio transients as a whole.

\section{Single-pulse search} \label{sec:search}

In this work, we searched the data for single pulses with a matched filter pipeline built using \presto. It differs slightly from the pipelines used to originally identify candidate sources in PALFA data. While the basic process is very similar, we describe it in full here for completeness.

First, the data were scanned for radio frequency interference (RFI) using \presto's \texttt{rfifind} utility.\footnote{When generating RFI masks, the \texttt{-time} option was set to 1.} The intervals and frequency channels identified by \texttt{rfifind} were subsequently excised. In a small number of cases, there was too much RFI for an observation to be used; these data are not included in this work.

Next, we performed a periodicity search on a timeseries dedispersed at the source DM with the \texttt{realfft} and \texttt{accelsearch} routines. For the sources with estimated periods, we also  folded the data at those periods to look for a significant folded pulse profile. J1929+1154 and J2010+3147 were detected in periodicity searches in approximately half of the follow-up observations.

We used \presto's \texttt{prepsubband} routine to dedisperse each observation at downsampling factors and trial DM spacing in accordance with the dedispersion plan described in \citealt{Lazarus+2015}. The use of many trial DMs allows for errors in discovery DM and, more importantly, helps distinguish astrophysical signals and radio frequency interference. For sources with a dispersion measure less than 212.8 pc cm$^{-3}$, we dedispersed at 101 trial DMs separated by 0.1 pc cm$^{-3}$, spanning a range $[\mathrm{DM}_s$ -- 5~pc cm$^{-3}, \mathrm{DM}_s$ + 5~pc cm$^{-3}$], with $\mathrm{DM}_s$ the source's dispersion measure. For sources with a dispersion measure of 212.8  pc cm$^{-3}$ or above, we dedispersed 35 trial DMs separated by 0.3 pc cm$^{-3}$, spanning a range $[\mathrm{DM}_s$ -- 5.1~pc cm$^{-3}$, $\mathrm{DM}_s$ + 5.1~pc cm$^{-3}]$, downsampling in time by a factor of 2.

\begin{figure}
\centering
\begin{minipage}{1\columnwidth}
  \centering
  \includegraphics[width=1\columnwidth]{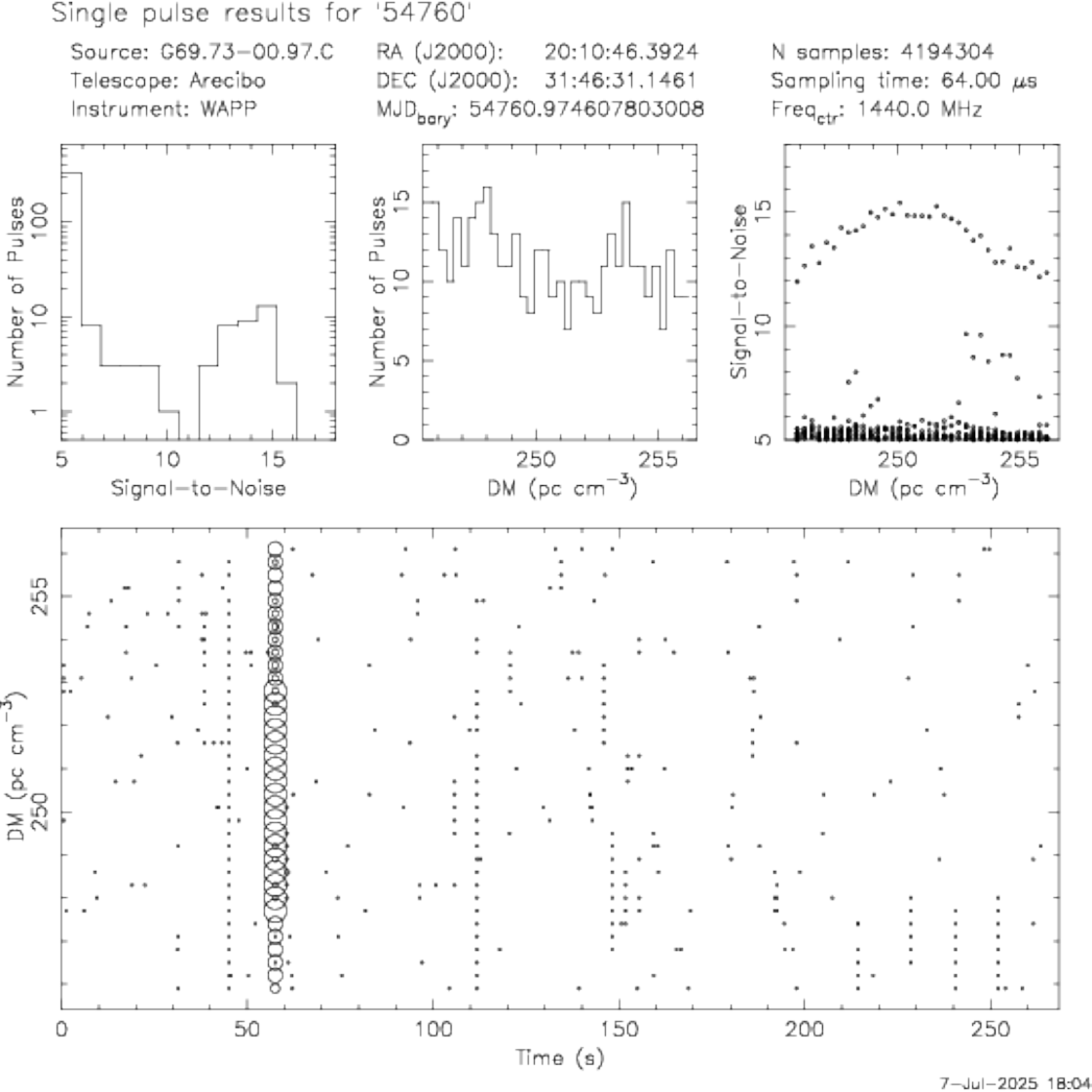}
\end{minipage}%
\begin{minipage}{1\columnwidth}
  \centering
  \includegraphics[width=1\columnwidth]{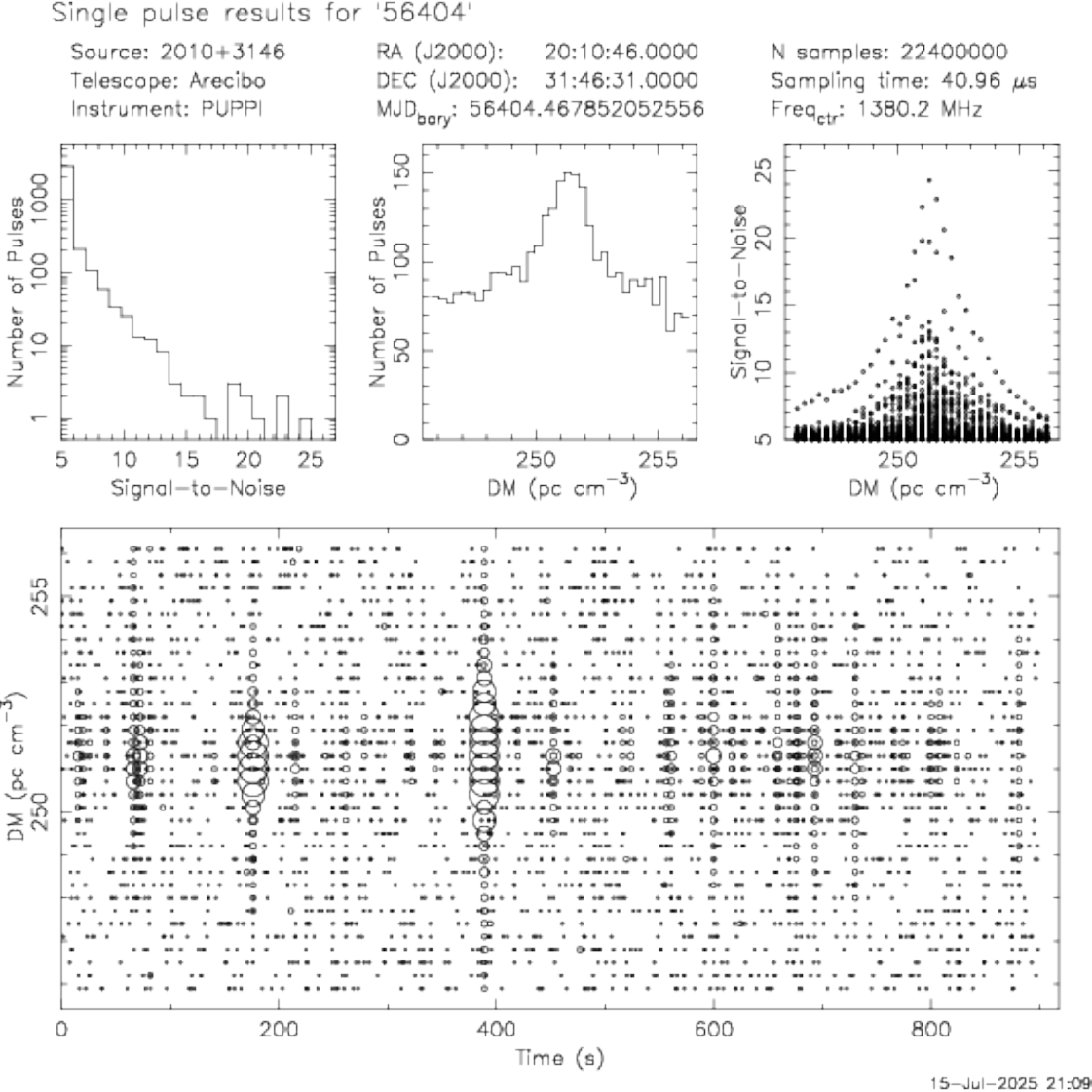}
\end{minipage}
\caption{Diagnostic single-pulse plots from the discovery observation of J2010+3147 (top), showing the original single pulse detected, and a follow-up timing observation of the same source (bottom). The former was taken using the ALFA/WAPP receiver/backend combination, while the latter was taken with the more sensitive LWB/PUPPI combination. The size of the circles in the bottom plot is proportion to the SNR of the pulse. This, along with the longer observation length, yielded 31 pulses, the strongest of which  clearly show the expected SNR/DM relation in the upper right subplot \citep{Cordes+McLaughlin2003}.}
\label{fig:j2010-sp-plots}
\end{figure}

We then used the matched filtering algorithm of \presto's \texttt{single\_pulse\_search.py} routine, which convolves each time series with a boxcar function, to search each for single pulses. The routine linearly detrends the data in each block of 1000 samples. Although the default maximum boxcar width is 30 samples, we increased the maximum width to 300 samples in the LBW/PUPPI observations, corresponding to a width of 12.289 ms.\footnote{This number was chosen based on experimentation, which found that shorter widths led to the nondetection of some real pulses.} This is approximately 12.2 ms, and the same width in time was used for the ALFA/Mock data. Candidate pulses with a signal-to-noise ratio (SNR) of 5 or above were recorded, and diagnostic plots were created for manual examination.\footnote{While we chose an SNR threshold of $6$ for the remainder of the single-pulse analyses, retaining candidates in the range $5\leq\mathrm{SNR}\leq6$ made it easier to identify the characteristic DM-SNR relation in single-pulse plots (see Section~3.3.3, \citealt{Cordes+McLaughlin2003}).} Examples of such plots are shown in Figure~\ref{fig:j2010-sp-plots}.

Single-pulse candidates were thinned further by searching for clusters of candidates detected at multiple DMs. The highest-SNR candidate was kept, and the others were discarded. At this stage, we now also discarded all candidates with SNRs below 6. All remaining candidates were inspected manually, using single-pulse profiles, the single-pulse diagnostic plots, and dynamic spectra; the latter were used to look for the characteristic time-frequency sweep of a dispersed astrophysical pulse.

\begin{table}
  \renewcommand\thetable{2}
  \centering
  \setlength{\tabcolsep}{1mm}
  \begin{tabular}{c c c c c c c}
  \hline
Source & $N_{\mathrm{obs}}$ & $T_{\mathrm{obs}}$ &$T_{\mathrm{obs},f}$ & $N_p$ & $N_f$ & $\mathcal{R}$ \\
& & (min) & (min) & & & (pulses min$^{-1}$) \\
\hline\hline
J0529+25 & 19 & 290.27 & 290.27 & 1 & 1 & $0.003^{+0.003}_{-0.003}$ \\
J0613+18 & 29 & 502.15 & 502.15 & 1 & 1 & $0.002^{+0.002}_{-0.002}$ \\
J0623+15 & 35 & 458.16 & 458.16 & 1 & 1 & $0.002^{+0.002}_{-0.002}$ \\
J0625+12 & 30 & 401.96 & 392.13 & 14 & 13 & $0.03^{+0.01}_{-0.01}$ \\
J1843+05 & 3 & 24.94 & 24.94 & 4 & 4 & $0.2^{+0.1}_{-0.1}$ \\
J1859+07 & 28 & 331.24 & 326.64 & 3 & 2 & $0.006^{+0.008}_{-0.005}$ \\
J1905+0413 & 28 & 319.61 & 315.01 & 141 & 135 & $0.43\pm0.04$ \\
J1906+0335 & 16 & 153.27 & 153.27 & 72 & 72 & $0.47\pm0.06$ \\
J1917+1142 & 21 & 289.27 & 284.65 & 54 & 51 & $0.18\pm0.03$ \\
J1924+10 & 3 & 24.94 & 20.34 & 11 & 9 & $0.4^{+0.2}_{-0.1}$ \\
J1929+1154 & 33 & 284.71 & 280.24 & 298 & 294 & $1.05\pm0.06$ \\
J2010+3147 & 31 & 324.65 & 320.18 & 865 & 864 & $2.7\pm0.1$ \\
\hline
\end{tabular}
{\footnotesize
  \caption{Results of the single-pulse search of Arecibo observations of the twelve sources. $N_{\mathrm{obs}}$ is the number of observations of the source, $T_{\mathrm{obs}}$ is the total observing time, $T_{\mathrm{obs,f}}$ is the observing time used for rate estimates, $N_p$ is the number of pulses, $N_f$ is the number of pulses used for rate estimates and fits, and $\mathcal{R}=N_f/T_{\mathrm{obs}}$ is the mean pulse rate. For all sources besides J0613+18 and J0623+15, the LBW/PUPPI observations are used to estimate a pulse rate. All quoted errors are $1\sigma$; we use $1\sigma$ limits calculate using the methods of \citealt{Gehrels1986} when $N_f<50$ and $1\sigma$ limits of $\sqrt{N_f}$ otherwise.}}
\end{table} \label{tab:sp-results}

\begin{table}
  \renewcommand\thetable{3}
  \centering
  \setlength{\tabcolsep}{1mm}
  \begin{tabular}{c c c c c}
  \hline
Source & $N_{\mathrm{obs}}$ & $T_{\mathrm{obs}}$ & $N_p$ & $\mathcal{R}$ \\
& & (min) & & (pulses min$^{-1}$) \\
\hline\hline
J0623+15 & 35 & 499.09 & 0 & $<0.0037$ \\
J0625+12 & 39 & 552.00 & 0 & $<0.0033$ \\
J1843+05 & 22 & 307.46 & 0 & $<0.0060$ \\
J1924+10 & 53 & 720.81 & 0 & $<0.0026$ \\
\hline
\end{tabular}
{\footnotesize
  \caption{Results of the single-pulse search of CHIME observations of the four sources. Pulse rates cited are $1\sigma$ upper limits (above the CHIME threshold) calculated according to the prescription of \citet{Gehrels1986}.}}
\end{table} \label{tab:chime-sp-results}

Table~\ref{tab:observation-details} in Appendix~\ref{sec:obs-details-appendix} lists the results of the single-pulse search for each Arecibo observation, and Table~\ref{tab:sp-results} provides a summary of all Arecibo observations of each source. Table~\ref{tab:chime-observation-details} and Table~\ref{tab:chime-sp-results} show the same information for the CHIME observations. Most of the sources were redetected in the Arecibo follow-up observations, with J1929+1154 proving to be the most active. J0613+18 and J0623+15 were not detected in follow-up observations and, along with J0625+12 and J1859+07, appear as fairly sporadic sources. By comparison, J1905+0413, J1906+0335, J1924+10 and J1929+1154 were extremely active. The discovery observation of J0529+25 has been lost, and so it could not be re-searched with this pipeline.

Finally, we estimated a rotation period for J1924+10 after detecting three pulses in the second observation and six pulses in the third. We again used \texttt{rrat\_period}, finding a period of 4.620 seconds.

\subsection{Sensitivity}

The sensitivity of a single-pulse search depends on a variety of instrumental factors. The minimal intrinsic peak flux density detectable is \citep{Deneva+2009}
\begin{equation}
S_{\mathrm{min}}=\frac{W}{W_i}\frac{\beta m\;\mathrm{SEFD}}{\sqrt{n_p\Delta fW}}
\end{equation}
with $m$ the SNR detection threshold, $\mathrm{SEFD}$ the system equivalent flux density,\footnote{Defined as the sum of the sky and receiver temperatures, divided by the gain.} $\beta\simeq1$ a digitization factor, $n_p$ the number of polarizations, $\Delta f$ the bandwidth, $W_i$ the intrinsic width of a pulse, and $W$ the observed width of a pulse once various broadening effects have been taken into account:
\begin{equation}\label{eqn:broadening}
W=\left(W_i^2+t_{\mathrm{samp}}^2+t_R^2+\Delta t_{\mathrm{DM,ch}}^2+\Delta t_{\mathrm{DM,err}}^2+\tau_{\mathrm{sc}}^2\right)^{1/2}
\end{equation}
where $t_{\mathrm{samp}}$ is the sampling time, $t_R\sim\Delta f^{-1}$ is the receiver filter response time, $\Delta t_{\mathrm{DM,ch}}$ is the broadening due to the dispersive smearing across a channel, $\Delta t_{\mathrm{DM,err}}$ is the broadening caused by the difference between the trial and true source DM, and $\tau_{\mathrm{sc}}$ is the broadening due to scattering. The number of polarizations, the bandwidth and the SEFD can all vary between receiver/backend combinations. Additionally, the second and fourth terms in Equation~\ref{eqn:broadening} depend on the particulars of the observing setup, because
\begin{equation}\label{eqn:smearing-and-scattering}
\Delta t_{\mathrm{DM,ch}}=8.3\;\mu\mathrm{s}\;\mathrm{DM}\Delta f_{\mathrm{ch}}/f^3,\quad\quad\tau_{\mathrm{sc}}\propto f^{-4}
\end{equation}
with $\Delta f_{\mathrm{ch}}$ the channel bandwidth measured in MHz and where the frequency is measured in GHz, and different receiver/backend combinations have different bandwidths and central frequencies. ALFA had a system temperature of $\sim30$ K and an SEFD of 2.8 Jy; the LBW had a system temperature of $\sim25$ K and an SEFD of 2.4 Jy. For a pulse width of $W\sim1$ ms, ignoring broadening effects, and $m=6$, we have nominal thresholds of $S_{\mathrm{min,AM}}\approx20.9$ mJy for the ALFA/Mock combination and $S_{\mathrm{min,LP}}\approx13.1$ mJy for the LBW/PUPPI combination, assuming we use LBW's effective bandwidth ($\Delta f_{\mathrm{eff}}=600$ MHz) instead of its nominal bandwidth ($\Delta  f=800$ MHz).

For our statistical analyses -- rate estimates and energy distributions -- of all sources with follow-up detections, we therefore only use the pulses from the LBW/PUPPI observations, and cite rates above the LBW/PUPPI threshold. For rate estimates of sources with nondetections in follow-up observations, we use all observations, and cite rates above the Mock/ALFA threshold.

For CHIME, using a nominal system temperature of $T_{\mathrm{sys}}\sim50$ K, a gain of $G\sim1$ K Jy$^{-1}$, and $n_p=2$ \citep{CHIME-Pulsar}, we find a sensitivity threshold of $S_{\mathrm{min},\mathrm{CHIME}}\approx335$ mJy for an SNR threshold $m=6$ and pulses of width $W=1$ ms. This does not take into account pulse broadening effects, which should be stronger in the CHIME band than at L-band, and so since $S_{\mathrm{min}}\propto\sqrt{W}$, this is likely an underestimate. Regardless, CHIME's decreased sensitivity means nondetections may be likely unless the sources have steep spectra.

We estimate the mean pulse rate for each source as the number of pulses, $N$, detected divided by the total observing time on the source, $T$. Assuming that the pulse rates obey Poisson statistics, the $1\sigma$ uncertainty on the rate is then $\sqrt{N}/T$.\footnote{This is of course an approximation. Assuming that single pulses are only emitted near one point in phase, there is a minimum possible separation in time between pulses. Additionally, many individual RRATs show inter-pulse wait time distributions inconsistent in other ways with Poisson statistics (see e.g. \citealt{McLaughlin+2009,Keane+2010,Keane+2011a,Palliyaguru+2011}). As such, our errors here should therefore be considered only rough estimates.} We use this for all sources with $N\geq50$; for sources with fewer pulses, we calculate $1\sigma$ upper and lower limits using the tables from \citet{Gehrels1986}. Observed pulse rates at Arecibo range from $\sim0.002$ to $\sim2.7$ pulses per minute. CHIME, with its lower sensitivity, resulted in no detections; we placed $1\sigma$ upper limits on pulse rates in the CHIME band again using the method of \citet{Gehrels1986}.

\subsection{Selection effects}

Beyond the flux density threshold established by choosing a minimum SNR of $m=6$, there are additional parts of single pulse parameter space that are undetectable due to selection effects. 
Pulses with widths less than the intrachannel dispersive smearing $\Delta t_{\mathrm{DM,ch}}$ listed in Equation~\ref{eqn:smearing-and-scattering} will be detectable only at reduced sensitivity; pulses with widths greater than the dispersive delay across the frequency band will appear consistent with a dispersion measure of zero, and will therefore be difficult to distinguish from RFI. Additionally, the set of matched filters used by the single-pulse search is finite and has some maximum width; pulses wider than this could still be detected if they are bright enough, but they will yield lower SNRs and may fall below the detection threshold.

In practice, the lower and higher width constraints are on the order of hundreds of microseconds and hundreds of milliseconds, respectively, and the single pulses we have detected typically fall in the middle of this range, far from either end. We also choose matched filters with widths up to tens of milliseconds to avoid the default cutoff made by the single-pulse search.

\subsection{Detrending and further analysis} \label{subsec:detrending}

To perform the statistical single pulse analyses presented in this paper, we need to extract additional information from each pulse, such as its equivalent width and fluence. This requires several steps. We first perform additional downsampling in time until the time series has been downsampled from its native time resolution by a factor of 8.

We then detrend the time series around each pulse. This requires several steps, including baseline subtraction. The simplest method would be to pick a single fixed-phase off-pulse region for each pulsar and use it for each single pulse, but there is significant pulse-to-pulse variation in pulse width and morphology. We therefore obtain an initial estimate of each pulse's width by fitting a Gaussian with a vertical offset to the region $|t_0-t|\leq L$, with $t_0$ the original estimated time of the pulse as determined by the single-pulse search. $L$ is chosen to be large enough to be much wider than the pulse but small enough to avoid significant wander in the baseline. The function fit is
\begin{equation}
f(t)=\frac{A_{\mathrm{est}}}{\sqrt{2\pi}\sigma_{\mathrm{est}}}\exp\left(-\frac{(t-\mu_{\mathrm{est}})^2}{2\sigma_{\mathrm{est}}^2}\right)+c
\end{equation}
with $A_{\mathrm{est}}$ the area under the Gaussian, $\mu_{\mathrm{est}}$ the center, and $\sigma_{\mathrm{est}}$ a characteristic width. We pick an initial off-pulse region satisfying
\begin{equation}
|\mu_{\mathrm{est}}-t|\geq 5\sigma_{\mathrm{est}},\quad\quad|\mu_{\mathrm{est}}-t|\leq l
\end{equation}
where $l$ is a length taken to be 512 bins for almost all sources, chosen to guarantee avoiding significant wander in the baseline.\footnote{For the case of J1929+1154, which shows features offset from the primary pulse, we replace the first inequality with $|\mu_{\mathrm{est}}-t|\leq 128$ bins.} We fit a line to this off-pulse region and use the slope and y-intercept to detrend the region $|\mu_{\mathrm{est}}-t|\leq L$. We then re-fit a Gaussian (with no vertical offset $c$) to the detrended data at $|\mu_{\mathrm{est}}-t|\leq L$, and obtain new values of $A$, $\mu$ and $\sigma$. All fits were performed using the least-squares \texttt{optimize.curve\_fit} routine from the \texttt{scipy}\footnote{\url{https://scipy.org/}} package, and were visually inspected.

The majority of detected pulses could have been detrended in a simpler fashion. The pulses are largely narrow, and there is rarely a notable change in the baseline over such small periods of time. Subtracting a constant, such as the median value over the region $|t_0-t|\leq L$, would have been equally effective in most cases. Indeed, this is an option available in \texttt{single\_pulse\_search.py}, which can speed up the search by a factor of $\sim2$ \citep{Ransom2011}. However, with only $\sim10^3$ pulses detected across the twelve sources, and with some pulses showing notable baseline changes, the speedup would not be worth the potential, albeit rare, loss of accuracy. Additional, more complex single-pulse detrending schemes have been used before, such as Gaussian process regression \citep{Shapiro-Albert+2018}. This appears unnecessary in this data set, as any notable baseline trends were quite linear.

Finally, quantities of interest for statistical analysis are computed. The SNR of a single pulse could be recomputed if the baseline trend was significant enough to add a systematic error to the results of \texttt{single\_pulse\_search.py}, though this does not appear to be the case for any of the pulses detected here. Another quantity of interest is the fluence $F$, which is computed by integrating the time series over the on-pulse region of each single pulse. From this, we can calculate the equivalent width $W_{\mathrm{eq}}$, defined as the width of a boxcar function with amplitude the same as the amplitude of the pulse. Errors for the fluence and equivalent width can also be estimated based on the off-pulse root-mean-squared noise fluctuations.

In Section~\ref{sec:pulse-energies}, we will fit distributions of pulse energies using fluence as a proxy. The traditional means of doing this is to equate energy with fluence, and, for the purpose of fitting, to normalize each pulse energy by the mean energy of an observation \citep{Burke-Spolaor+2012,Mickaliger+2018} or a subblock of an observation \citep{Ritchings1976,Biggs1992}. We do not do this here -- and instead normalize by the mean energy over \textit{all} observations of a source -- because the sources are quite sporadic, and per-observation energy means would be subject to errors due to small-number statistics, particularly for J1905+0413, J1906+0335 and J1917+1142. Our approach likely introduces errors due to variations in telescope sensitivity, but it is still presumably preferable to the alterative.

\subsection{Periodicity searches and single-pulse rates}

\begin{figure*}
    \centering
    \includegraphics[width=1.9\columnwidth]{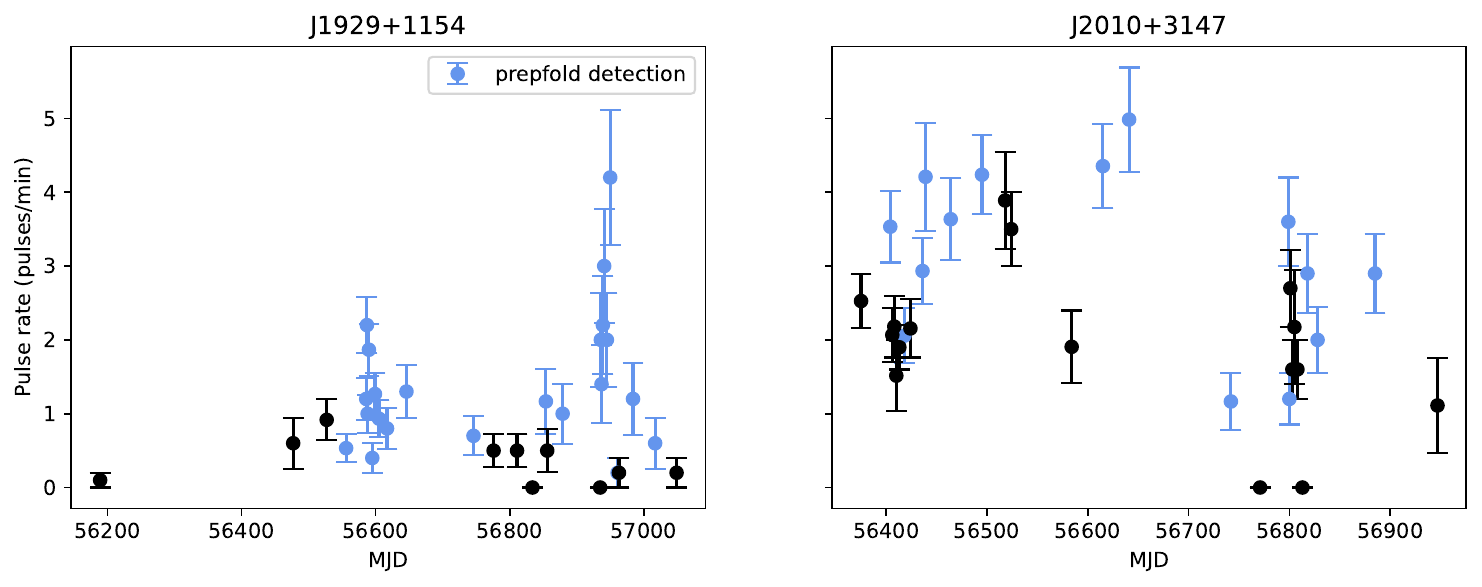}
    \caption{Measured pulse rates from the J1929+1154 (left) and J2010+3147 (right) follow-up observations, with errors corresponding to a Poisson process ($\sigma_{\mathcal{R}}=\sqrt{N_p}/T$), color-coded depending on whether or not a periodicity search was successful. As expected, the periodicity detections come from observations with higher pulse rates.}
    \label{fig:prepfold-single-pulse-comparison}
\end{figure*}

Although only J1929+1154 and J2010+3147 were ever detected in periodicity searches, there are unsurprising relationships between those detections and pulse rates. Figure~\ref{fig:prepfold-single-pulse-comparison} shows the distribution of the two subsets of single-pulse rates from these sources: those corresponding to detections using \texttt{prepfold} and those corresponding to non-detections. As expected, the periodicity search detections are largely from observations with the highest measured pulse rates.

\section{Timing} \label{sec:timing}

We attempted to obtain phase-connected timing solutions for all sources with sufficient number of pulses spanning an adequate number of epochs. As only small numbers of pulses were detected from some of the sources, this was only possible for J1905+0413, J1906+0335, J1917+1142, J1929+1154 and J2010+3147.

\begin{figure*}
    \centering
    \includegraphics[width=1.9\columnwidth]{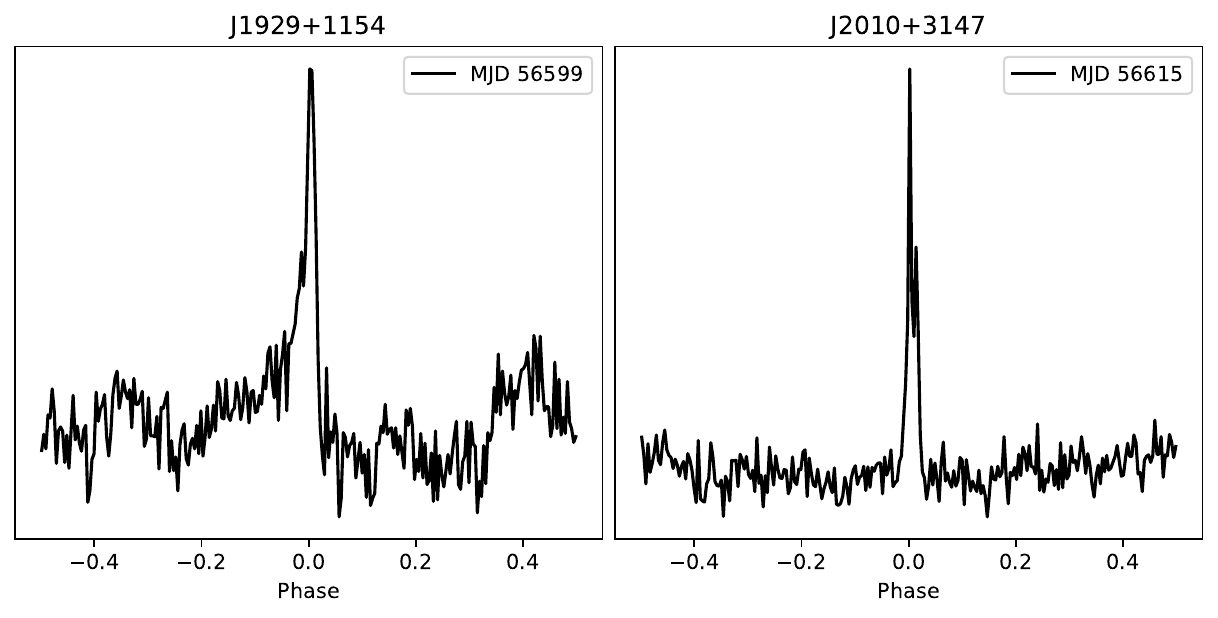}
    \caption{The composite template profiles for J1929+1154 and J2010+3147. These were chosen because they were the profiles with the highest signal-to-noise ratios and minimal interference, and were not smoothed when used to generate TOAs.}
    \label{fig:folded-profiles}
\end{figure*}

Times of arrival (TOAs) were generated using the FFTFIT frequency-domain algorithm \citep{Taylor1992}, with correlates a single pulse or a folded pulse profile with a template. For all five sources, we obtained TOAs from single pulses. A template was obtained by adding together all of the single-pulse profiles for a given source. We did so by fitting a Gaussian to each single pulse, then aligning the centers of the Gaussians; almost all single pulses were well-approximated by a single component.\footnote{Although some pulses from J1929+1154 displayed multiple components, as noted in Section~\ref{subsec:j1929-substructure}, there was too much variation in the separation of the components to make a two-peak template effective without inducing additional systematic uncertainty. For simplicity, we use the single-Gaussian template for all pulses.} For J1929+1154 and J2010+3147, we also obtained TOAs from folded profiles using \presto's \texttt{get\_TOAs.py} routine, choosing the template to be the folded profile from the epoch with the highest folded profile SNR and minimal interference, as computed by \texttt{prepfold}; between 1 and 4 TOAs were generated per epoch, depending on how much of the observation the source was detectable in and its brightness. We timed these two pulsars twice, once using single-pulse TOAs and once using folded profile TOAs, and the other three pulsars only through single pulses. The composite profiles used for templates for J1929+1154 and J2010+3147 are shown in Figure~\ref{fig:folded-profiles}. 

The nominal uncertainty in a single-pulse TOA should be of order \citep{Lorimer+Kramer2004}
\begin{equation}
\sigma_{\mathrm{TOA}}\simeq\frac{W}{\mathrm{SNR}}
\end{equation}
with $W$ the pulse width and $\mathrm{SNR}$ the single-pulse signal-to-noise ratio. Each single pulse TOA was assigned an uncertainty accordingly. However, RRAT single pulses exhibit significant pulse-to-pulse variation in phase (``pulse jitter'') and shape \citep{Keane+2011a}. This adds to radiometer noise to increase the root mean square of the pulse residuals; the total TOA error comes from the error due to pulse jitter, $\sigma_{\mathrm{jit}}$ added in quadrature with the error due to radiometer noise, $\sigma_{\mathrm{rad}}$:
\begin{equation}
\sigma_{\mathrm{tot}}^2=\sigma_{\mathrm{rad}}^2+\sigma_{\mathrm{jit}}^2
\end{equation}
The timing baselines for these sources range from 476 days (J1917+1142) to 1240 days (J1905+0413). These should be sufficient to fit for spin frequency $f$, right ascension, declination and  spin frequency time derivative $\dot{f}$. There is no indication that any of these sources are in binary systems or necessitate more complicated timing models.

For several sources, TOAs are sparse, making manual phase connection more difficult. Motivated by this, we attempted timing solutions using the Algorithmic Pulsar Timer for Binaries\footnote{https://github.com/Jackson-D-Taylor/APT} (APTB; \citealt{APTB}), which has proven successful with sparse TOAs. Designed to produce timing solutions for binary pulsars, APTB uses a brute-force depth-first algorithm to time pulsars algorithmically. APTB is also capable of timing isolated pulsars, and because it improves over its predecessor, the Algorithmic Pulsar Timer (APT; \citealt{APT}), in several other respects, we chose to use APTB instead of APT. Where APT was unable to find a solution, we attempted timing solutions by hand using the \texttt{PINT}\footnote{\url{https://github.com/nanograv/PINT}} package \citep{PINT}, which APTB also uses.

The sources have narrow folded profiles and even narrower single pulses. This means that even good solutions should have large reduced chi-squared ($\chi_r^2$) values, which APTB typically interprets as implying that a trial solution is incorrect. While this can be ameliorated by raising the threshold $\chi_r^2$ value that tells APTB to prune a branch, doing so also ensures APTB will do a poor job of pruning incorrect solutions. In order to solve this problem, we artificially increase the computed uncertainty on each single-pulse TOA by one order of magnitude. This is essentially inserting an error fraction, or ``EFAC", a multiplicative white noise component.

\begin{figure*}
\centering
\includegraphics[width=1.9\columnwidth]{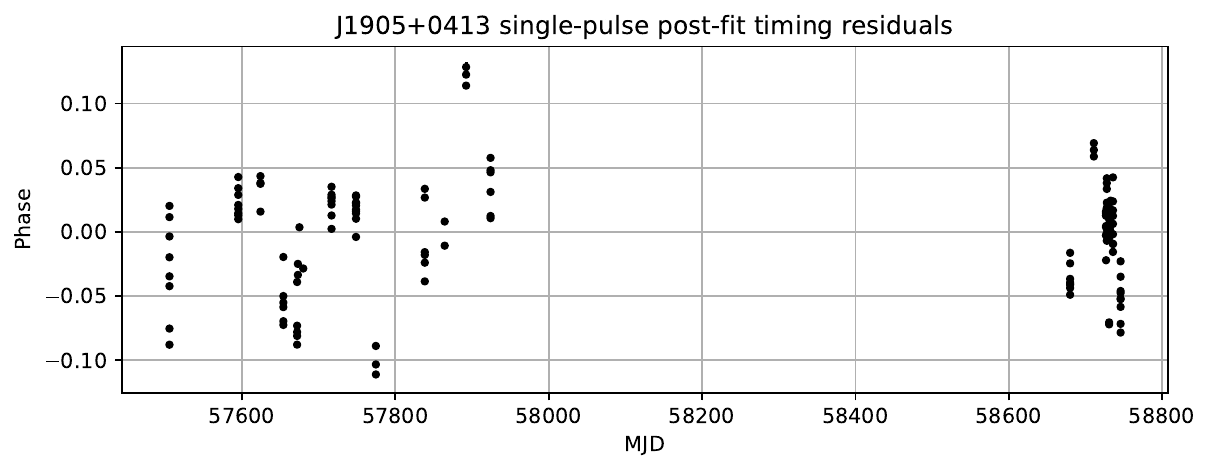} \label{fig:j1905-sp-solution}
\includegraphics[width=1.9\columnwidth]{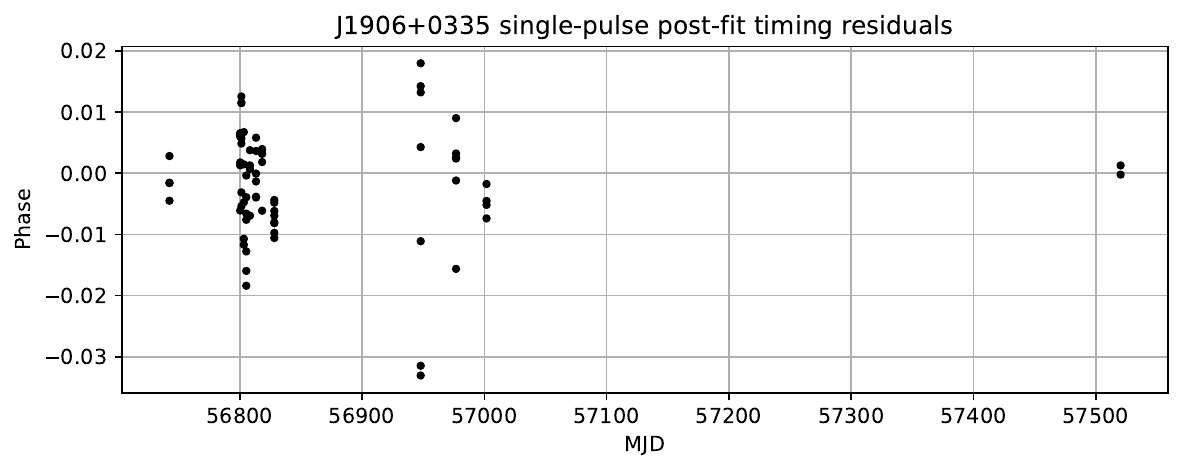} \label{fig:j1906-sp-solution}
\includegraphics[width=1.9\columnwidth]{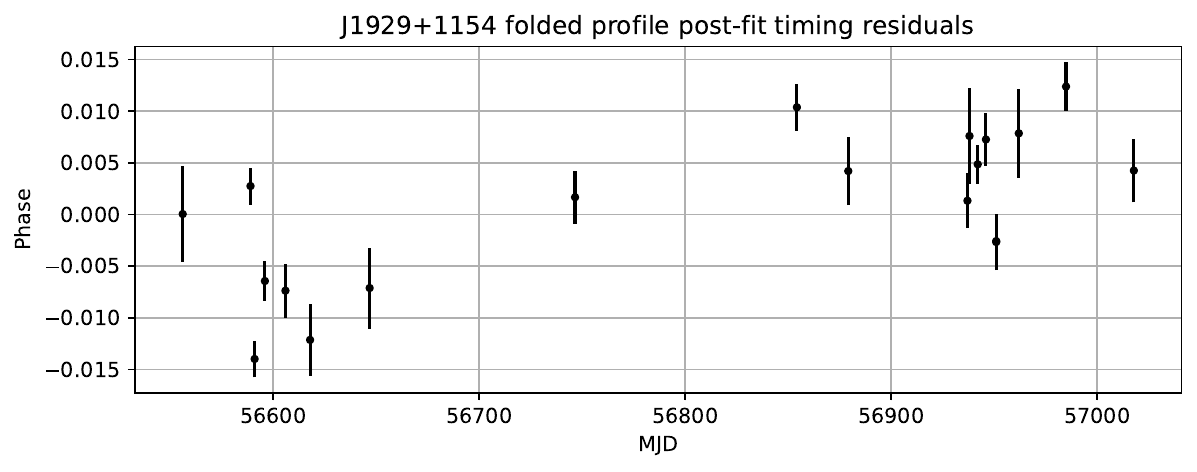} \label{fig:j1929-folded-solution}
\caption{Single-pulse timing residuals for J1905+0413 (top) and J1906+0335 (middle), and the folded profile solution for J1929+1154 (bottom).} \label{fig:residuals-one}
\end{figure*}

\begin{figure*}
\centering
\includegraphics[width=1.9\columnwidth]{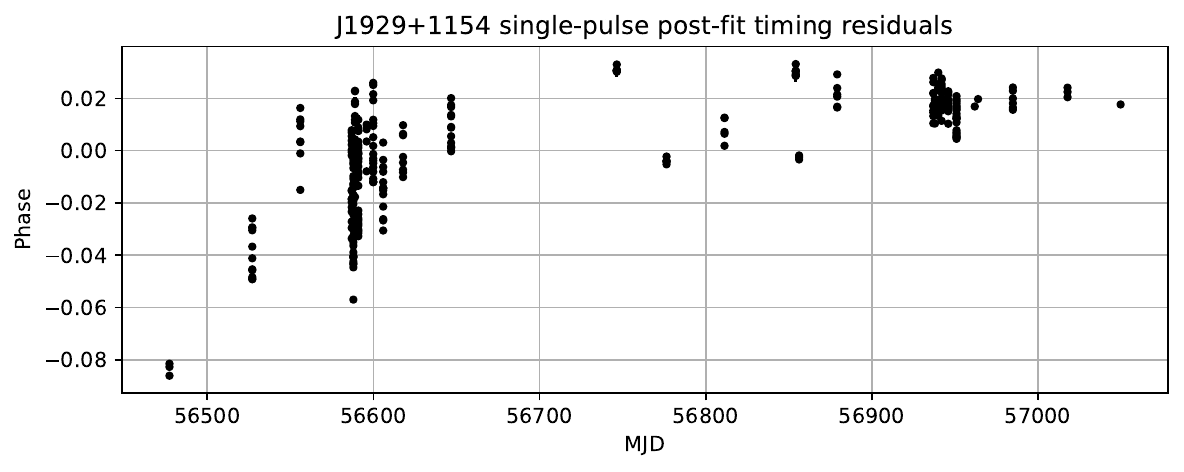} \label{fig:j1929-sp-solution}
\includegraphics[width=1.9\columnwidth]{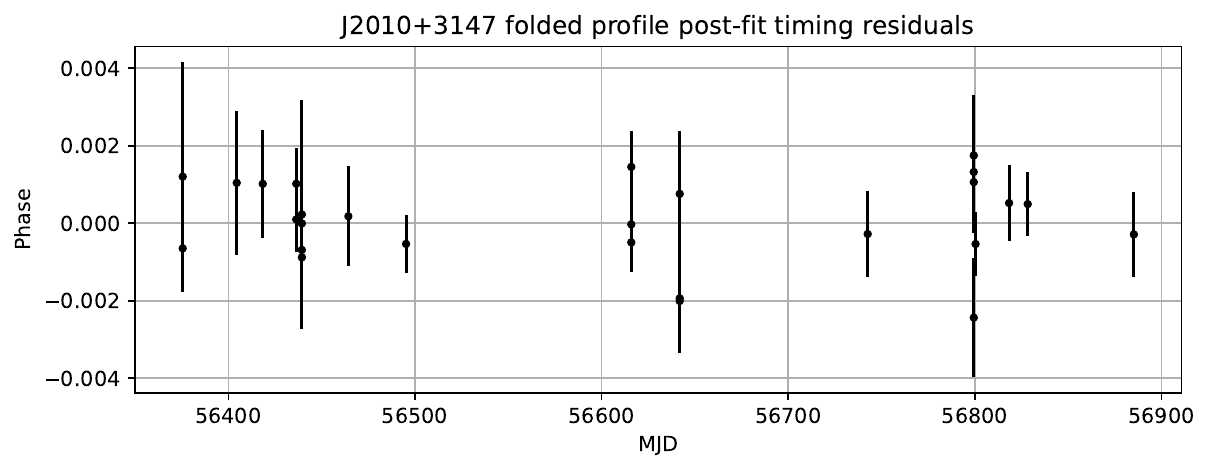} \label{fig:j2010-folded-solution}
\includegraphics[width=1.9\columnwidth]{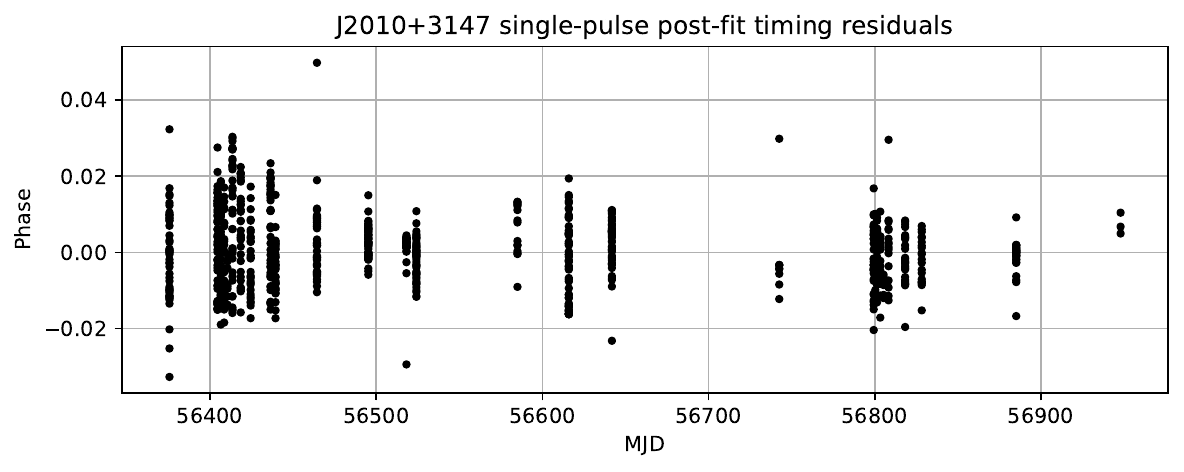} \label{fig:j2010-sp-solution}
\caption{Single-pulse solutions for J1929+1154 (top) and J2010+3147 (bottom), and the folded profile solution for J2010+3147 (middle).} \label{fig:residuals-two}
\end{figure*}

Timing solutions are listed in Table~\ref{tab:timing}. As is typical, we list spin period $P$ and period derivative $\dot{P}$, with uncertainties propagated in quadrature, instead of $f$ and $\dot{f}$. We successfully obtained solutions through single-pulse TOAs for J1905+0413 and J1906+0335. We obtained a solution through folded profile TOAs for J1929+1154 and J2010+3147, and then used those as starting points to obtain solutions based on the single-pulse TOAs. Plots of timing residuals are shown in Figure~\ref{fig:residuals-one} and Figure~\ref{fig:residuals-two}.

\begin{table*}
  \renewcommand\thetable{4}
  \centering
  \setlength{\tabcolsep}{1mm}
  \begin{tabular}{l c c c c c c}
  \hline
  Source & J1905+0413 & J1906+0335 & J1929+1154 & J1929+1154 & J2010+3147 & J2010+3147 \\
  & & & (folded) & (single-pulse) & (folded) & (single-pulse) \\
  \hline\hline
  Epoch (MJD) & 50000 & 50000 & 50000 & 50000 & 56735 & 56375 \\
  DM (pc cm$^{-3}$) & 183 & 212 & 80 & 80 & 251 & 251 \\
  Timespan (yr) & 3.40 & 2.13 & 1.26 & 1.57 & 1.40 & 1.57 \\
  TOAs & 135 & 72 & 19 & 293 & 27 & 864 \\
  Right ascension & 19:05:15.76(3) & 19:06:51.51(7) & 19:29:18.3(1) & 19:29:18.3(1) & 20:10:35.99(2) & 20:10:36.737(3)\\
  Declination & 4:13:10(1) & 3:35:41.5(8) & 11:54:38(4) & 11:54:38(4) & 31:47:32.6(4) & 31:47:33.91(4) \\
  Period, $P$ (s) & 0.894072669(1) & 1.296418546(7) & 3.2172793(1) & 3.2172793(1) & 1.5514534482(2) & 1.55145344721(3) \\
  Period derivative, $\dot{P}$ & $2.16(2)\times10^{-16}$ & $8.8(1)\times10^{-16}$ & $4(2)\times10^{-16}$ & $4(2)\times10^{-16}$ & $1.7729(1)\times10^{-13}$ & $1.77337(1)\times10^{-13}$ \\
  Residual RMS, $\sigma_{\mathrm{rms}}$ (ms) & 29.9 & 11.1 & 23.3 & 71.1 & 1.39 & 14.3 \\
  Characteristic age (yr) & $6.6\times10^7$ & $2.3\times10^7$ & $1.1\times10^8$ & $1.1\times10^8$ & $1.4\times10^5$ & $1.4\times10^5$ \\
  Magnetic field strength (G) & $4.5\times10^{11}$ & $1.1\times10^{12}$ & $1.3\times10^{12}$ & $1.3\times10^{12}$ & $1.7\times10^{13}$ & $1.7\times10^{13}$ \\
  Spin-down luminosity (erg s$^{-1}$) & $1.2\times10^{31}$ & $1.6\times10^{31}$ & $5.7\times10^{29}$ & $5.7\times10^{29}$ & $1.9\times10^{33}$ & $1.9\times10^{33}$ \\\hline
  \end{tabular}
{\footnotesize
  \caption{Timing parameters for the pulsars with sufficient detections to be successfully timed by hand or with APTB. The errors in parentheses are the 1$\sigma$ errors reported by PINT. Although $P$ and $\dot{P}$ are listed here, we fit for spin frequency$f$ and its time derivative $\dot{f}$. When computing the surface magnetic field strength and spin-down luminosity, we use a nominal moment of inertia of $I=10^{45}$ g cm$^2$ and a radius $R=10$ km and assume that the pulsar is spinning down solely due to magnetic dipole radiation and that the spin and magnetic axes are orthogonal.}}
\end{table*} \label{tab:timing}

\begin{figure*}
\centering
\includegraphics[width=1.9\columnwidth]{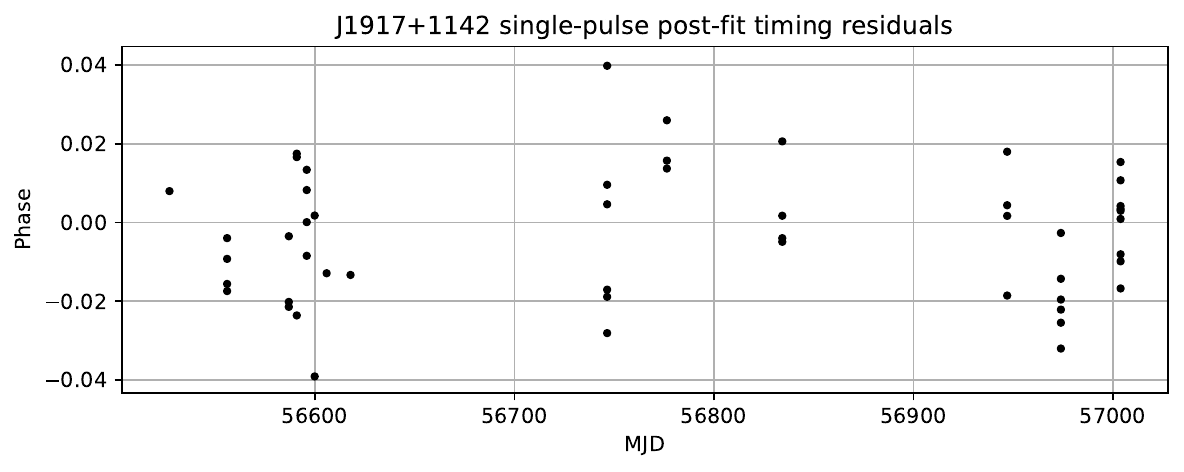} \label{fig:J1917+11-sp-solution}
\caption{The single-pulse solution for J1917+1142.}
\end{figure*}

\begin{table*}
  \renewcommand\thetable{5}
  \centering
  \setlength{\tabcolsep}{1mm}
  \begin{tabular}{l c}
  \hline
  Source & J1917+1142 \\
  & \\
  \hline\hline
  Epoch (MJD) & 57003 \\
  DM (pc cm$^{-3}$) & 319 \\
  Timespan (yr) & 1.30 \\
  TOAs & 51 \\
  Right ascension & 19:17:07.0(1) \\
  Declination & 11:42:51(3) \\
  Period, $P$ (s) & 1.187916045(1) \\
  Period derivative, $\dot{P}$ & $2.820(25)\times10^{-14}$ \\
  Residual RMS, $\sigma_{\mathrm{rms}}$ (ms) & 19.2 \\
  Characteristic age (yr) & $6.7\times10^5$ \\
  Magnetic field strength (G) & $5.9\times10^{12}$ \\
  Spin-down luminosity (erg s$^{-1}$) & $6.6\times10^{32}$ \\\hline
  \end{tabular}
{\footnotesize
  \caption{Timing parameters for J1917+1142, after refitting the DRACULA solution with PINT. The errors in parentheses are the 1$\sigma$ errors reported by PINT. As with the other pulsars, we fit for spin frequency$f$ and its time derivative $\dot{f}$, and when computing the surface magnetic field strength and spin-down luminosity, we use a nominal moment of inertia of $I=10^{45}$ g cm$^2$ and a radius $R=10$ km and assume that the pulsar is spinning down solely due to magnetic dipole radiation and that the spin and magnetic axes are orthogonal.}}
\end{table*} \label{tab:j1917-timing}

\subsection{J1905+0413} \label{j1905-timing}

The gap of approximately two years made it difficult to phase connect the TOAs from J1905+0413 manually. However, APTB successfully obtained a solution. The root mean square (RMS) of the residuals is approximately 3\% in phase, the largest in the sample.

\subsection{J1906+0335} \label{j1906-timing}

Similarly, APTB found a solution for J1906+0335. The RMS is significantly better. However, the vast majority of the TOAs fall into a window of roughly one year, making it difficult to measured $\dot{f}$ (and, by extension, $\dot{P}$) with much accuracy.

\subsection{J1917+1142} \label{j1917-timing}

APTB was not able to solve J1917+1142, nor were human timers. It was possible to manually achieve phase connection over short spans, but not on timescales longer than months to a year. We were, however, successfully able to obtain a solution with the similar, older software DRACULA\footnote{https://github.com/pfreire163/Dracula} \citep{DRACULA}, which is based on the pulsar timing software \tempo\footnote{https://tempo.sourceforge.net/} \citep{Tempo2}. For consistency with the other solutions, we subsequently refit it with PINT, finding no significant changes to the fit.

It is unclear why DRACULA was successful but APTB was not. It is notable that DRACULA produced many alternate solutions that featured incorrect coordinates, which is likely due to the short timing baseline ($\sim1.3$ years) combined with a small number of TOAs, but it is unclear why this would have affected APTB and not DRACULA. The timing residuals are displayed in Figure~\ref{fig:J1917+11-sp-solution} and the fit and derived parameters are listed in Table~\ref{tab:j1917-timing}.

\subsection{J1929+1154} \label{j1929-timing}

APTB could not solve J1929+1154, regardless of whether it was given TOAs from single pulses or from folded profiles. In the single-pulse case, the large number of TOAs, coupled with intrinsic pulse-to-pulse scatter, meant that even a good solution would have a comparatively large chi-squared. Using APTB therefore required overriding default settings and allowing it to stay on a branch even when it found a high $\chi^2$. This significantly increased the size of the parameter space explored by the code, but did not yield any results.

Similarly, APTB did not find a solution using folded profile TOAs. This may have been due in part to the faintness of the pulsar; it was difficult to justify generating more than one TOA per epoch. However, a human was able to manually find a solution. This was subsequently re-fit using the TOAs from single pulses, after removing the TOA from MJD 54907. Neither solution was able to constrain $\dot{f}$ with any significance.\footnote{To check the $\dot{f}$ fit, we used \presto's \texttt{rrat\_period} function to estimate a period at each epoch, then fit a line to calculate $\dot{P}$. The fit yielded $\dot{P}=4(10)\times10^{-13}$, which poorly constrains $\dot{P}$ and does not allow us to draw any significant conclusions.} Adding $\ddot{f}$ to the timing model did not result in an improvement. We note moderate covariances between the position and $\dot{f}$.

\subsection{J2010+3147} \label{j2010-timing}

APTB was able to time J2010+3147 through its folded profiles. The resulting solution is, by RMS, the best we were able to generate. As with J1929+1154, APTB was unable to find a single-pulse solution on its own because of difficulties with branches with large $\chi^2$ values; we did fit the single-pulse TOAs starting from the folded profile solution. The two solutions are quite close in $f$ (with $P$ differing by $\sim1$ ns) and $\dot{f}$ (with $\dot{P}$ differing by $\sim10^{-17}$) but differ notably in coordinates (close to $\sim1.2$ arcseconds in right ascension and $\sim1.3$ arcseconds in declination).

\section{Single-pulse morphology} \label{sec:morphology}

\begin{figure*}
\centering
\includegraphics[width=1.9\columnwidth]{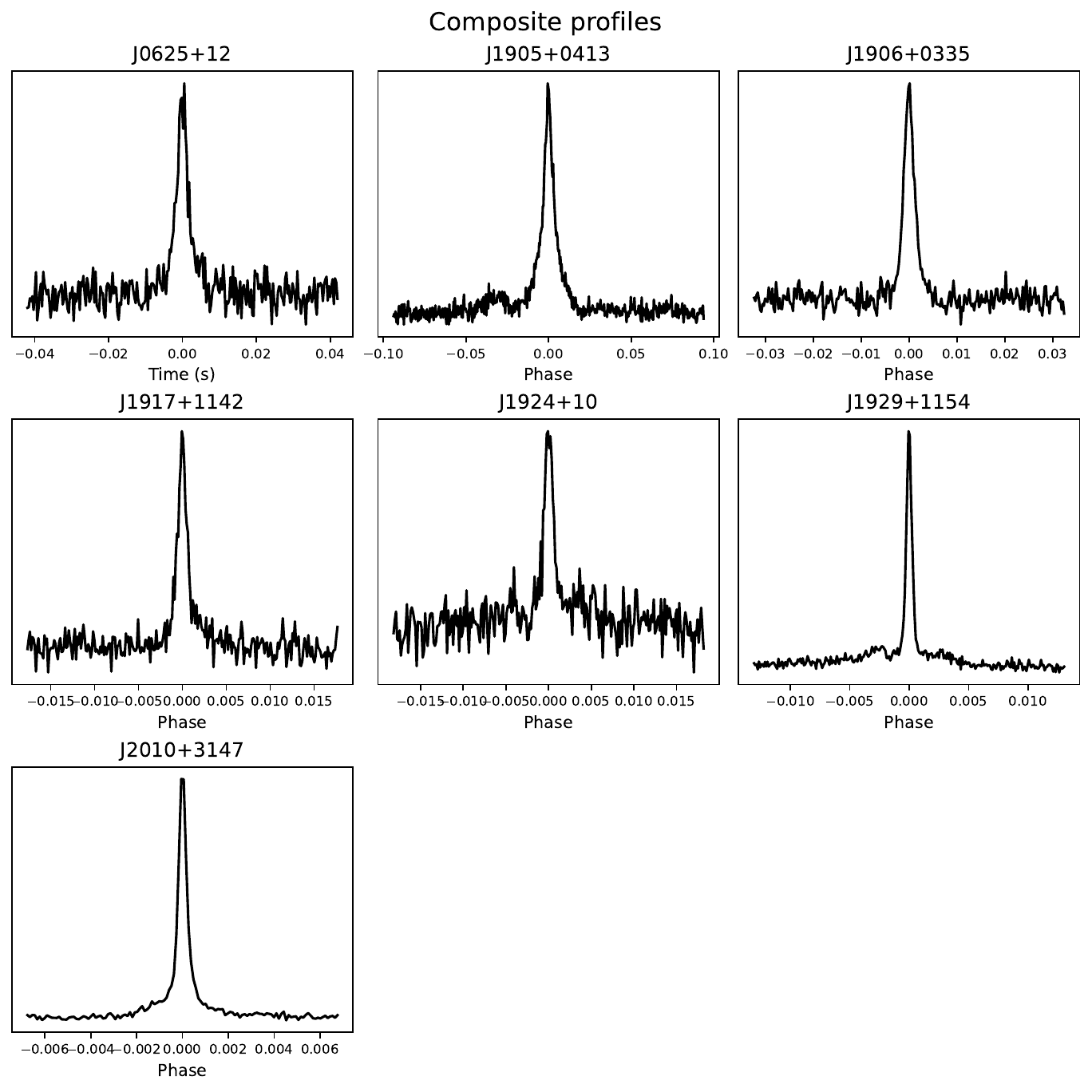} \label{fig:composite-profiles}
\caption{Composite profiles for the seven sources which have adequate numbers of single pulses to create a representative profile. Pulses were aligned by fitting Gaussians to each pulse and aligning the centers of the Gaussians. This results in composite profiles with much higher signal-to-noise ratios that would be achieved by simply folding an entire observation at the pulse period, since no rotations with just noise are included. The horizontal axes are units of rotation phase, with the exception of J0625+12, for which no period is known.}
\end{figure*}

\begin{figure*}
\centering
\includegraphics[width=1.9\columnwidth]{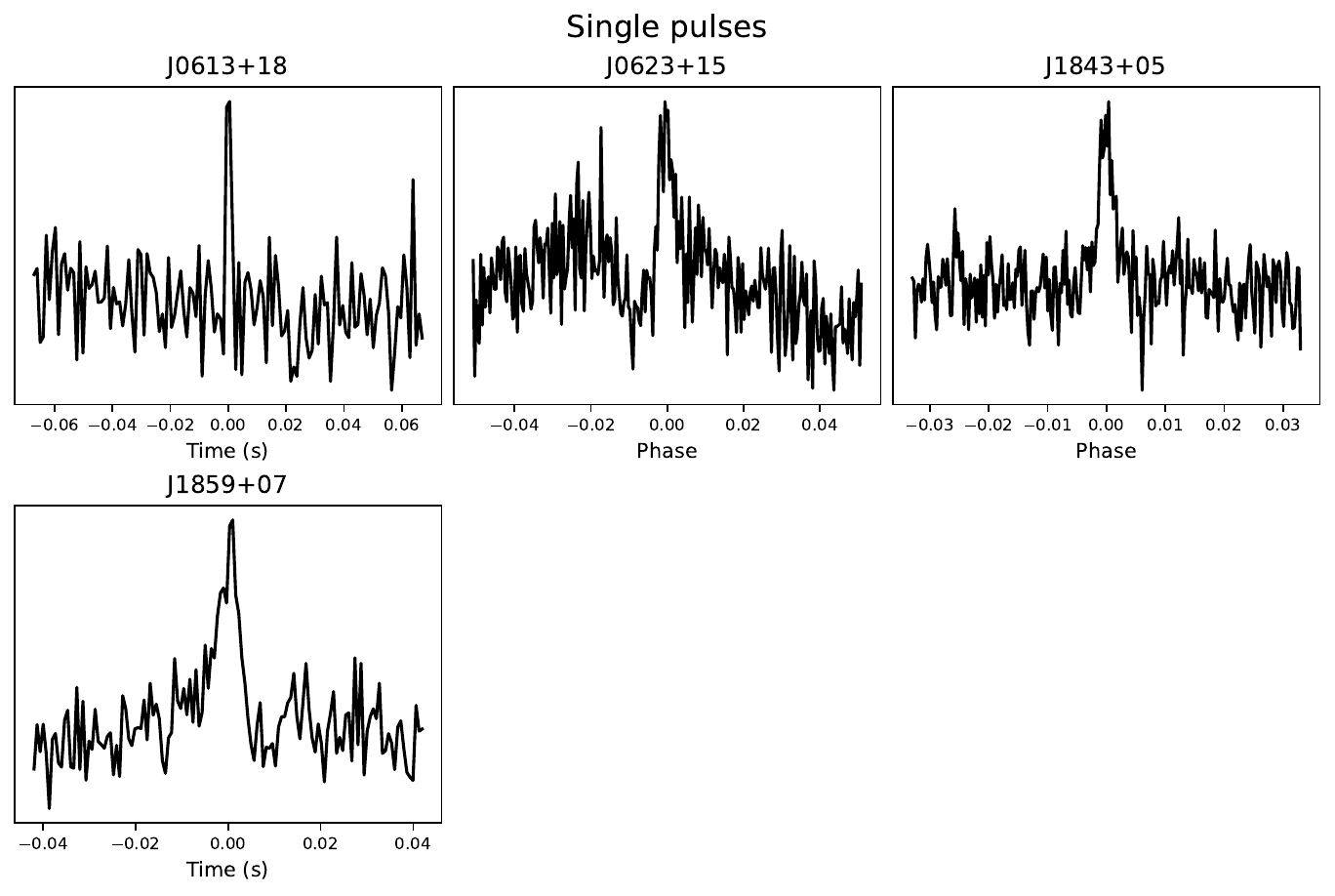} \label{fig:single-profiles}
\caption{The strongest single pulses for each of four of the remaining five sources. The J1859+07 pulse came from the observation on MJD 57505; all others came from the corresponding discovery observation. The period from the Galactic Plane Pulsar Snapshot Survey (GPPS; see Section~\ref{subsec:GPPS}) was used to compute the phase range for J0623+15.}
\end{figure*}

We obtained composite profiles in the same manner as the templates used in Section~\ref{sec:timing} for all sources with a suitable number of single pulses. Composite profiles are shown in Figure~\ref{fig:composite-profiles}, and the strongest single pulses from each of four of the remaining five sources are shown in Figure~\ref{fig:single-profiles}. The discovery data for J0529+25 has been lost and is therefore unavailable for plotting.

\subsection{J0613+18} \label{subsec:J0613+18-profile}

The discovery observation of J0613+18 shows a single narrow, broadband Gaussian pulse. These features are  characteristic of nonrepeating FRBs \citep{CHIME2021,Pleunis+2021,Curtin+2024}, which is unsurprising given the high DM and the nondetection of the source in all follow-up observations.

\subsection{J1905+0413}

Pulses from J1905+0413 are well-represented by two components. \citealt{GPPS2} derived a complicated composite profile for J1905+0413, featuring two major components separated by approximately 6\% in phase. The second, brighter, component features substructure, including a small pair of peaks at its leading edge spanning only a few percent in phase. It appears to be these two narrow peaks that comprise our composite profile; the rest of the structure is not detectable in our observations, although we do see a possible component leading the main pulse by approximately 4\% in phase.

\subsection{J1924+10}

The pulses from J1924+10 are again quite narrow compared to the \citealt{GPPS2} composite profile. This is unsurprising, however, as the profile's relative broadness is due to jitter amongst the single pulses.

\subsection{Substructure from J1929+1154} \label{subsec:j1929-substructure}

\begin{figure}
    \centering
    \includegraphics[width=0.9\columnwidth]{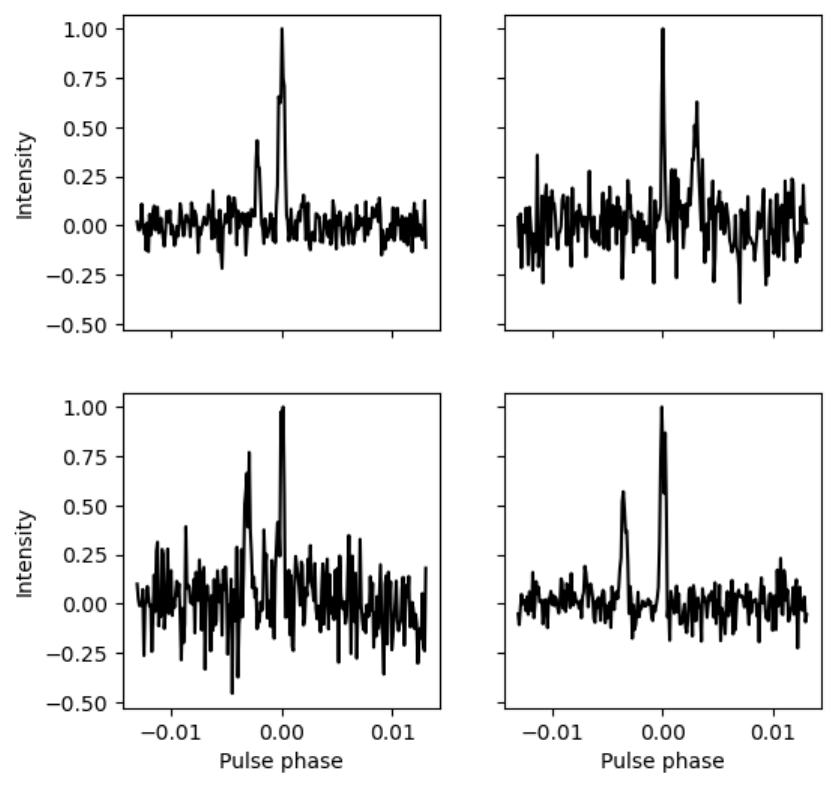}
    \caption{Four examples of the minority of pulses from J1929+1154 that exhibit substructure. In most multi-component pulses, a weaker component precedes the stronger, but there are some exceptions, such as the pulse shown in the upper right panel.}
    \label{fig:j1929-microstructure}
\end{figure}

A minority of J1929+1154's pulses feature substructure, manifesting as two or three distinct spikes. A selection of these are shown in Figure~\ref{fig:j1929-microstructure}. The weaker component usually, but not always, leads the primary peak.

The separations of these spikes are typically roughly $\sim10$ ms, or approximately $\sim3\times10^{-3}$ times the spin period. This delay is notable; a selection of pulsars, magnetars, RRATs and FRBs have been found to display quasiperiodic features with a quasiperiodicity closely tied to spin period \citep{Kramer+2024}:
\begin{equation}
P_{\mu}=(0.94\pm0.04)\;\left(\frac{P}{\mathrm{s}}\right)^{(0.97\pm0.05)}\;\mathrm{ms}
\end{equation}
Given J1929+1154's spin period of $P\approx3.217$ seconds, the relation predicts that any quasiperiodic features would have quasiperiodicity $P_{\mu}\approx3$ ms, which is only a factor of three shorter than the inter-peak separation seen in J1929+1154. It is therefore possible that the features seen in J1929+1154 are examples of this quasiperiodicity. However, the vast majority of the multi-component pulses from J1929+1154 feature only two components, while many compact objects exhibiting quasiperiodicity show far more. Therefore, it is difficult to have any confidence that this phenomenon is actually due to the same mechanism as quasiperiodicity.

\section{Pulse energy distributions} \label{sec:pulse-energies}

Distributions of RRAT single-pulse energies can be complex. Various emission models provide physical justifications for log-normal, Gaussian and power-law fits \citep{Cairns+2003a,Cairns+2003b}. In practice, Gaussians and power laws appear to be disfavored for pulsars and RRATs, with log-normal distributions providing significantly better fits \citep{Burke-Spolaor+2012}. Other works have found that more complex distributions, such as power laws with exponential cutoffs, can even improve upon log-normal fits \citep{Mickaliger+2018}. Broken power laws have been used for certain individual RRATs \citep{Chen+2022} but not yet, to our knowledge, to large portions of the RRAT population.

There are five sources in our sample which have a sufficient number of pulses for reasonable fits: J1905+0413, J1906+0335, J1917+1142, J1929+1154 and J2010+3147. J1906+0335 and J1917+1142 yielded significantly fewer pulses than the RRATs and pulsars used in prior studies, but still enough to perform fits. The distributions of J1905+0413, J1917+1142 and J1929+1154 are bimodal, with a small number of high-energy pulses, while J1906+0335 and J2010+3147 are unimodal. Based on this and previous studies in the literature, we are primarily interested in fitting two distributions: a log-normal distribution, with mean $\mu$ and standard deviation $\sigma$
\begin{equation}
P_{\mathrm{LN}}(E|\mu,\sigma) = \frac{1}{\sqrt{2\pi}E\sigma}\exp\left[-\frac{(\ln(E) - \mu)^2}{2\sigma^2}\right]
\end{equation}
and a power-law fit with exponential cutoff
\begin{equation}
P_{\mathrm{PE}}(E|b,c) = \frac{c}{\Gamma(b+1)}\frac{(cE)^b}{e^{cE}}
\end{equation}
with $\Gamma(x)$ the gamma function, $c$  the inverse of a characteristic energy scale, and $b$  a parameter describing the sharpness of the power-law falloff. The second  distribution is just a reparameterization of the gamma distribution.

The histograms of the fluences of the pulses indicate that neither power laws nor broken power laws fit our data well. We therefore do not perform those fits. For the RRATs with bimodal distributions, we only fit the lower-energy set, which constitutes the vast majority of the pulses. All energies are normalized by dividing by the mean of all pulses to be fitted in the source, as discussed in Section~\ref{subsec:detrending}.

\subsection{Least-squares histogram fitting}

A traditional method of fitting energy distributions involves binning the pulses by energy into a histogram. The center energies of the bins and the number of pulses in each bin are then used as inputs in a least-squares routine to fit a desired distribution \citep{Burke-Spolaor+2012,Mickaliger+2018}. In the case of a power law distribution, for example, this procedure can be reduced to fitting a straight line, assuming the energies and pulse numbers are taken to be in log-log space.

Different distributions can be compared to one another by computing a reduced chi-squared statistic $\chi_{\mathrm{red}}^2$ for each distribution $P(E)$:
\begin{equation}
\chi_{\mathrm{red}}^2=\frac{1}{N_f-m}\sum_{i=1}^n\frac{(N_i-p_i)^2}{\sigma_i^2}
\end{equation}
with $N_f$ the number of pulses, $m$ the number of model parameters, and $n$ the number of bins. $N_i$, $p_i$ and $\sigma_i^2$ are the number of pulses, the distribution's prediction for the number of pulses, and the variance in the number of pulses, in bin $i$. The distribution with the lowest $\chi_{\mathrm{red}}^2$ is selected.

Since the distribution is a probability density function but the $N_i$ are counts, $p_i$ is taken to be
\begin{equation}
p_i=N_f\int_{E_{\mathrm{low},i}}^{E_{\mathrm{high},i}}P(E)\;\mathrm{d}E
\end{equation}
with $E_{\mathrm{low},i}$ and $E_{\mathrm{high},i}$ the low and high edges of energy bin $i$.

We perform a least-squares fit for each of our five sources using the recipe above. We choose $n=10$ logarithmically-spaced bins for the two sparser sources, J1906+0335 and J1917+1142, and $n=20$ logarithmically-spaced bins for the more active three, J1905+0413, J1929+1154 and J2010+3147. We set $\sigma_i^2=N_i$, though we note that others (see e.g. \citealt{Burke-Spolaor+2012}) have chosen $\sigma_i^2=p_i$; this avoids cases where $N_i=0$, which would lead to ill-defined values of $\chi_{\mathrm{red}}^2$. Such a case does not show up in our distributions.

The histogram method is appealing. It is quick to implement and widely used in the literature, and the procedure for comparing models is simple and intuitive. However, there are significant problems with this approach, particularly in cases of small sample size (see \citealt{Goldstein+2004,Clauset+2009} for a discussion of these in the context of power law fitting). Biases can be introduced, proper error estimation is difficult, and the choice of binning scheme can introduce systematic errors. Additionally, fits are not always properly normalized, unless appropriate care is taken.

Our sample could be particularly prone to many of these issues because the most active sources in it produced relatively few pulses compared to RRATs from earlier studies. We explore this possibility in Appendix~\ref{sec:energy-fitting-appendix}, and find that it performs poorer than the method described in Section~\ref{subsec:bayesian-fitting}. We therefore use the histogram method only as a baseline for comparison with our other results, given its ubiquity in the literature.

\subsection{Bayesian MCMC fitting} \label{subsec:bayesian-fitting}

In recent years, better approaches have been applied to the task of fitting energy distributions. Some previous works have applied maximum likelihood estimators to distributions from fast radio bursts and Crab giant pulses \citep{Crawford+1970,Oronsaye+2015,Meyers+2017,James+2019,Crawford+2023}. These are particularly useful for distributions whose likelihoods can be maximized analytically, such as power law and log-normal distributions. After computing best-fit parameters by maximizing the likelihood, uncertainties can then be obtained from the diagonal entries of the Fisher matrix.

We choose to go beyond this and perform Markov Chain Monte Carlo (MCMC) sampling. This allows us to derive proper posteriors and explore any possible covariances between parameters. For each distribution we wish to fit, we first invoke Bayes' theorem:
\begin{equation}
\mathcal{P}(\boldsymbol{\theta}|\boldsymbol{E})\propto\mathcal{L}(\boldsymbol{E}|\boldsymbol{\theta})\mathcal{P}(\boldsymbol{\theta})
\end{equation}
with $\boldsymbol{\theta}=\{\theta_1,\theta_2,...\}$ a set of parameters and $\boldsymbol{E}=\{E_1,E_2,...\}$ the observed pulse energies. $\mathcal{P}(\boldsymbol{\theta}|\boldsymbol{E})$ is the posterior distribution, $\mathcal{L}(\boldsymbol{E}|\boldsymbol{\theta})$ is the likelihood, and $\mathcal{P}(\boldsymbol{\theta})$ is the prior. Appropriately sampling the parameter space of $\boldsymbol{\theta}$ allows us to numerically obtain posterior distributions. In the context of RRAT single-pulse statistics, similar approaches have been taken at least once before, to model the SNR distributions of 155 single pulses from the RRAT J0628+0909 \citep{Hsu+2023}\footnote{The details of the method used in \citealt{Hsu+2023} are unclear from the text, and their method may differ from ours.} and to compute distribution parameters and nulling fractions from the Thousand Pulsar Array program \citep{Keith+2026}.

Assuming that the energy of each pulse is independent of the others, the likelihood of a set of pulse energies is equal to the product of the likelihoods of each individual pulse energy:
\begin{equation}
\mathcal{L}(\boldsymbol{E}|\boldsymbol{\theta})=\prod_{i=1}^{N_f}\mathcal{L}(E_i|\boldsymbol{\theta})=\prod_{i=1}^{N_f}P(E_i|\boldsymbol{\theta})
\end{equation}
We choose uniform priors for $\mu$ ($-3\leq\mu\leq3$) and $b$ ($-1\leq b\leq6$) and log-uniform priors for $\sigma$ ($-2\leq\ln\sigma\leq2$) and $c$ ($-1\leq\ln c\leq2$). We performed the sampling with the \texttt{emcee}\footnote{\url{https://github.com/dfm/emcee}} package \citep{emcee}, obtaining Bayesian means and confidence intervals for all fitted parameters. In Appendix~\ref{sec:energy-fitting-appendix}, we compare the performance of the procedure to the performance of the least squares histogram recipe, and find that the MCMC approach is generally more accurate than the standard least squares method.

The logical recipe for model selection in the Bayesian method is to compute a Bayes factor. The Bayes factor of model $\mathcal{M}_i$ relative to model $\mathcal{M}_j$ is
\begin{equation}
\mathcal{B}_{ij}=\frac{\mathcal{P}(\mathcal{M}_i|\boldsymbol{E})}{\mathcal{P}(\mathcal{M}_j|\boldsymbol{E})}
\end{equation}
where $\mathcal{P}(\mathcal{M}_i|\boldsymbol{E})$ is the marginal likelihood of model $i$, assuming the distributions are believed to be equally likely \textit{a priori}. The task is then to compute each marginal likelihood. For numerical stability, in practice we in fact compute the natural logarithms of the likelihoods, and from there $\ln\mathcal{B}_{ij}$.\footnote{In this work, there are only two models of interest. For clarity, in the rest of this paper, we use $\mathcal{B}$ to denote the Bayes factor of the log-normal distribution compared to the power law/exponential distribution.}

A simple approximation applicable to MCMC sampling is the harmonic mean estimator \citep{Newton+1994}, which computes the marginal likelihood from the likelihoods of individual samples:
\begin{equation}
\mathcal{P}(\mathcal{M}_i|\boldsymbol{E})\approx\left[\frac{1}{N}\sum_{n=1}^N\frac{1}{\mathcal{L}_i(\boldsymbol{E}|\boldsymbol{\theta}_n)}\right]^{-1}
\end{equation}
where $\boldsymbol{\theta}_n$ is the set of parameter values from sample $n$ and the sum is taken over a set of $N$ samples. The harmonic mean estimator can be computed directly from the MCMC samples and is quite easy to implement. However, its variance may be extremely large or even infinite, depending upon the tails of the priors and posteriors \citep{Neal1994}, which is of course problematic. Therefore, naive harmonic mean estimates of Bayes factors should be viewed cautiously, and avoided when possible. Fortunately, modifications of the estimator have been developed to solve the variance problem. Many introduce a new density $\varphi(\boldsymbol{\theta})$, chosen to modify the tails to avoid infinite variance \citep{Gelfand+Dey1994}:
\begin{equation}
\mathcal{P}(\mathcal{M}_i|\boldsymbol{E})\approx\left[\frac{1}{N}\sum_{n=1}^N\frac{\varphi(\boldsymbol{\theta}_n)}{\mathcal{L}_i(\boldsymbol{E}|\boldsymbol{\theta}_n)\mathcal{P}(\boldsymbol{\theta}_n)}\right]^{-1}
\end{equation}
The recipe for choosing $\varphi(\boldsymbol{\theta})$ is not always clear. We apply the learned harmonic mean estimator, implemented in the \texttt{harmonic}\footnote{\url{https://github.com/astro-informatics/harmonic}} package \citep{harmonic}, which uses a machine learning technique to choose an appropriate density. \texttt{harmonic} has the advantage of working particularly well with \texttt{emcee} and other affine invariant sampling packages.\footnote{For each set of energies, we did also compute $\ln\mathcal{B}$ using the plain harmonic mean estimator. In all cases, our results differed from the \texttt{harmonic} results by no more than $\Delta\ln\mathcal{B}\simeq1$, indicating that there was no catastrophic failure of the naive recipe. However, the success of the harmonic mean estimator in this small number of cases should not overcome the significant statistical concerns about its usage in general!}

We note that a Bayes factor should be interpreted slightly differently than a comparison of two reduced chi-squared values. A reduced chi-squared describes a distribution with specific values of its parameters chosen; a Bayes factor describes a distribution without fixing its parameters. In this sense, a Bayes factor is better suited to answer the question of whether a RRAT's pulses are more properly described by a log-normal energy distribution or a power law/exponential distribution.

\subsection{Results}

\begin{table*}
  \renewcommand\thetable{6}
  \centering
  \setlength{\tabcolsep}{1mm}
  \begin{tabular}{c | c c c c c c | c c c c c}
  \hline
  
& \multicolumn{6}{c|}{Least-squares} & \multicolumn{5}{c}{MCMC} \\
Source & $\mu$ & $\ln\sigma$ & $\chi_{\mathrm{red,LN}}^2$ & $b$ & $\ln c$ & $\chi_{\mathrm{red,PE}}^2$ & $\mu$ & $\ln\sigma$ & $b$ & $\ln c$ & $\ln\mathcal{B}$ \\
\hline\hline
J1905+0413 & $-0.42^{+0.09}_{-0.09}$ & $-0.37^{+0.12}_{-0.12}$ & $2.02$ & $1.80^{+0.65}_{-0.65}$ & $1.34^{+0.30}_{-0.30}$ & $4.58$ & $-0.31^{+0.07}_{-0.07}$ & $-0.24^{+0.06}_{-0.06}$ & $0.76^{+0.20}_{-0.19}$ & $0.56^{+0.13}_{-0.13}$ & $6.30$ \\
J1906+0335 & $-0.04^{+0.12}_{-0.12}$ & $-0.55^{+0.20}_{-0.20}$ & $1.32$ & $2.68^{+1.23}_{-1.23}$ & $1.29^{+0.39}_{-0.39}$ & $1.14$ & $-0.11^{+0.06}_{-0.06}$ & $-0.71^{+0.09}_{-0.08}$ & $3.68^{+0.78}_{-0.71}$ & $1.54^{+0.16}_{-0.17}$ & $-3.38$ \\
J1917+1142 & $-0.20^{+0.05}_{-0.05}$ & $-0.55^{+0.07}_{-0.07}$ & $0.44$ & $2.90^{+0.38}_{-0.38}$ & $1.51^{+0.12}_{-0.12}$ & $0.65$ & $-0.19^{+0.08}_{-0.09}$ & $-0.49^{+0.10}_{-0.10}$ & $1.82^{+0.56}_{-0.49}$ & $1.04^{+0.20}_{-0.21}$ & $1.04$ \\
J1929+1154 & $-0.29^{+0.03}_{-0.03}$ & $-0.39^{+0.04}_{-0.04}$ & $0.91$ & $2.00^{+0.27}_{-0.27}$ & $1.30^{+0.12}_{-0.12}$ & $2.90$ & $-0.28^{+0.04}_{-0.04}$ & $-0.32^{+0.04}_{-0.04}$ & $0.94^{+0.15}_{-0.14}$ & $0.66^{+0.08}_{-0.09}$ & $20.89$ \\
J2010+3147 & $-0.20^{+0.02}_{-0.02}$ & $-0.67^{+0.03}_{-0.03}$ & $1.70$ & $3.71^{+0.34}_{-0.34}$ & $1.69^{+0.08}_{-0.08}$ & $4.89$ & $-0.17^{+0.02}_{-0.02}$ & $-0.59^{+0.02}_{-0.02}$ & $2.17^{+0.15}_{-0.14}$ & $1.15^{+0.05}_{-0.05}$ & $58.43$ \\
\hline
\end{tabular}
{\footnotesize
  \caption{The results of the energy fits for the five sources with a sufficient number of pulses. In general, the log-normal fits for both the least squares and MCMC methods are consistent, while the power-law/exponential fits differ significantly. $\ln\mathcal{B}$ is the logarithmic Bayes factor for preferring the log-normal model over the power-law/exponential model. All errors are $1\sigma$.}}
\end{table*} \label{tab:energy-fits}

\begin{figure*}
\centering
\begin{minipage}{1.9\columnwidth}
  \centering
  \includegraphics[width=1\columnwidth]{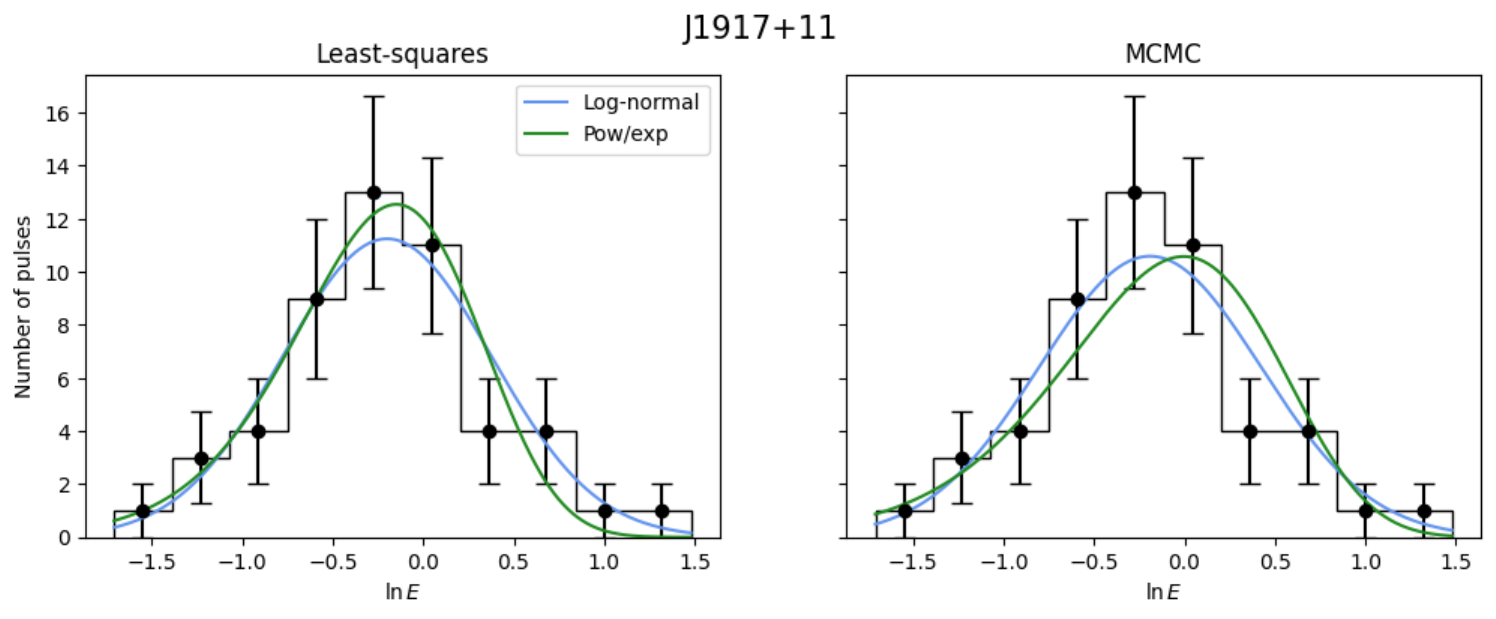}
  \label{fig:j1917-energy-fits}
\end{minipage}%
\begin{minipage}{1.9\columnwidth}
  \centering
  \includegraphics[width=1\columnwidth]{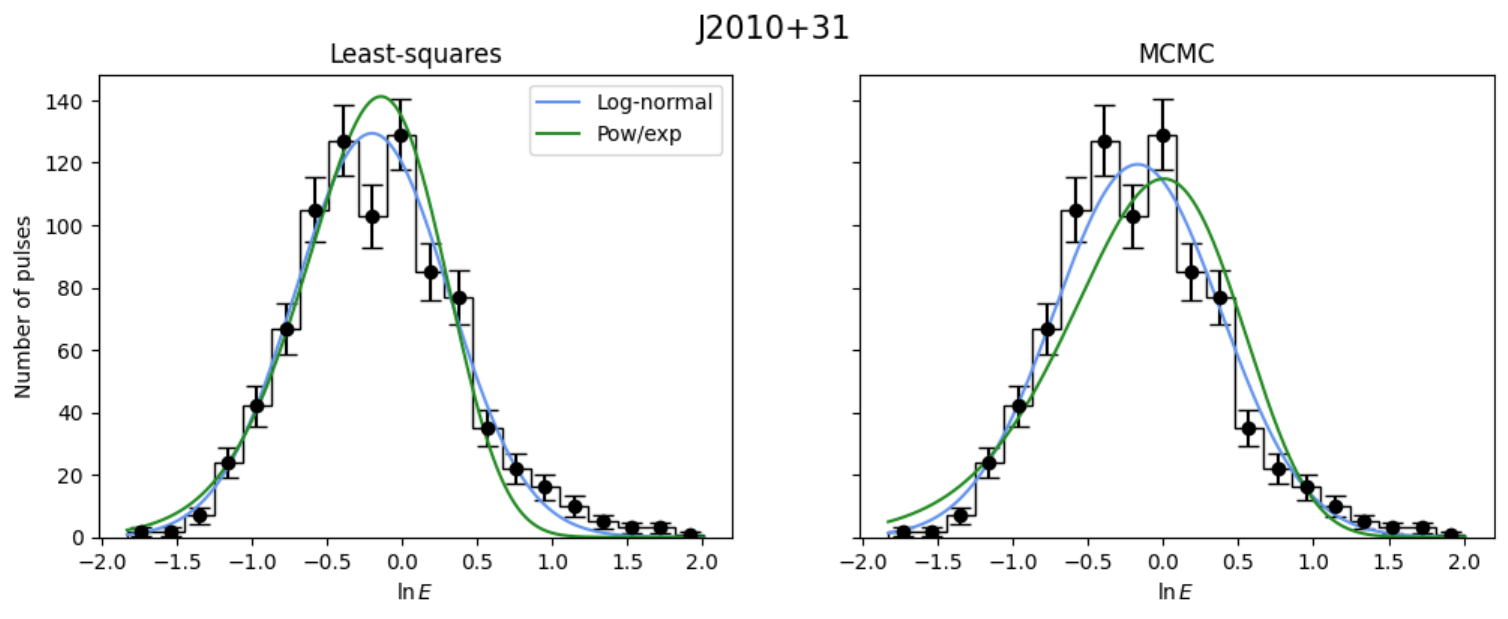}
  \label{fig:j2010-energy-fits}
\end{minipage}
\caption{Pulse energy distributions for J1917+1142 (top) and J2010+3147 (bottom). The left plots also show the fits from the least-squares method; the right plots show the fits from the MCMC method. The error bars on bins are derived assuming Poisson statistics.}
\end{figure*}

\begin{figure*}
\centering
\begin{minipage}{1.9\columnwidth}
  \centering
  \includegraphics[width=1\columnwidth]{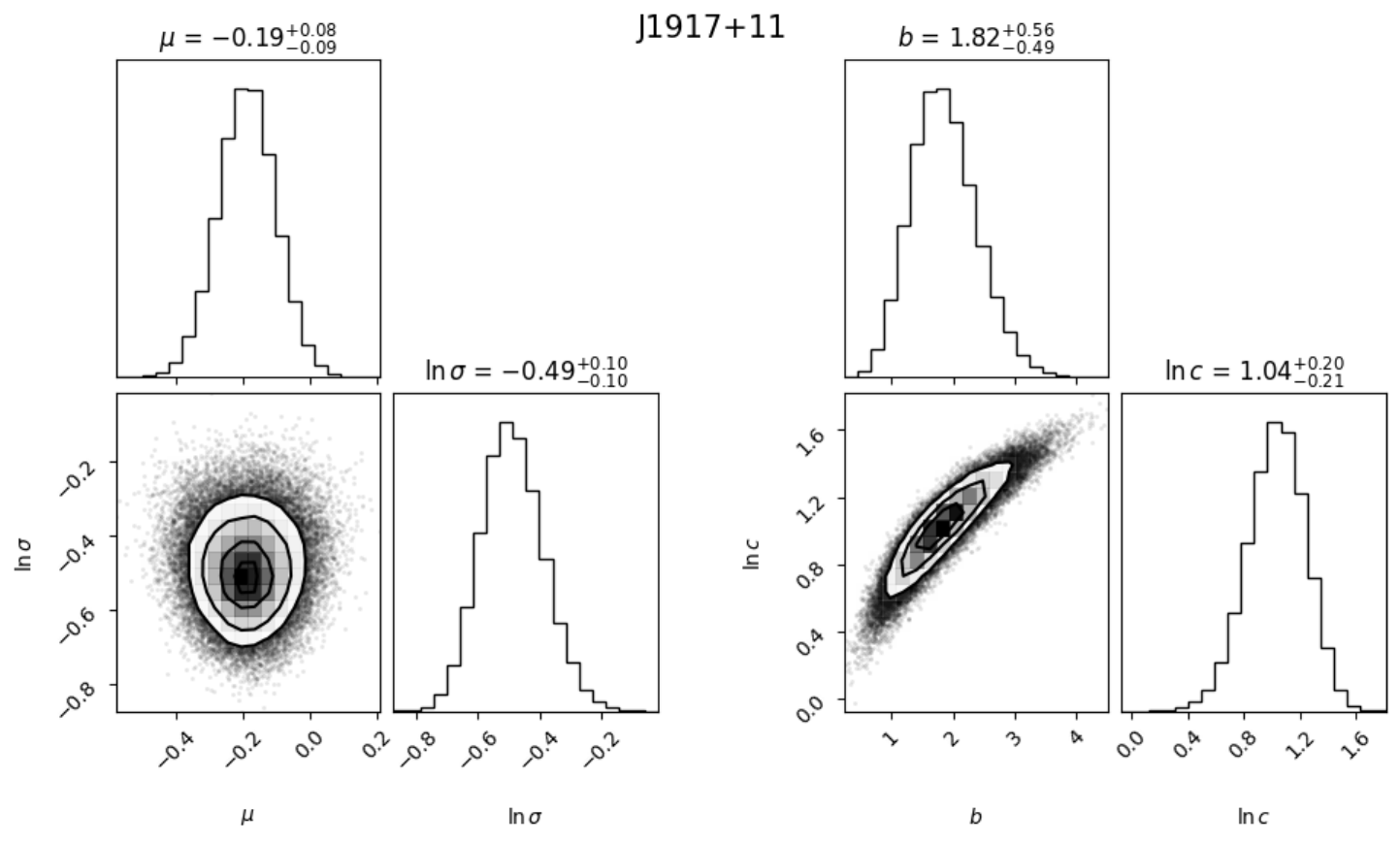}
  \label{fig:j1917-corner-plots}
\end{minipage}%
\begin{minipage}{1.9\columnwidth}
  \centering
  \includegraphics[width=1\columnwidth]{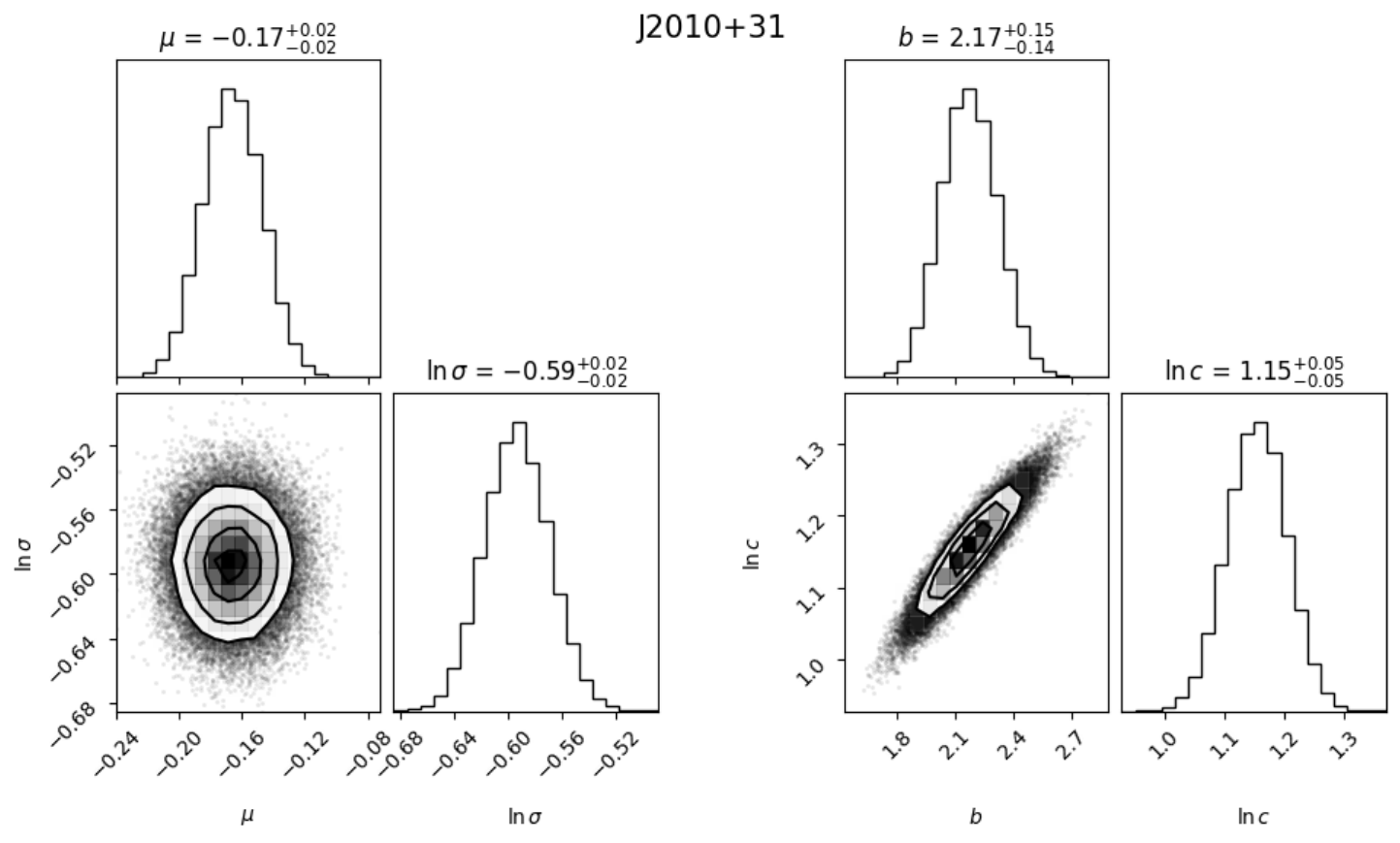}
  \label{fig:j2010-corner-plots}
\end{minipage}
\caption{Corner plots from the MCMC fits for J1917+1142 (top) and J2010+3147 (bottom). There is a strong covariance between $b$ and $\ln c$, but minimal covariance between $\mu$ and $\ln\sigma$. See Section~\ref{sec:pulse-energies} for more details on the distributions and fitting method.}
\end{figure*}

Table~\ref{tab:energy-fits} shows the results of the fits and the associated model selection statistics. Figure~\ref{fig:j1917-energy-fits} shows plots of the fits for the source with the fewest pulses of the five (J1917+1142) and the source with the most pulses (J2010+3147), and Figure~\ref{fig:j1917-corner-plots} shows corner plots from the MCMC fits for those two sources. The posteriors show that, in general, there is a strong covariance between $b$ and $c$ when fitting the power-law/exponential distribution.

In four of the pulsars, the log-normal model is preferred over the power law/exponential model, with logarithmic Bayes factors between $\ln\mathcal{B}=1.04$ and $\ln\mathcal{B}=58.43$. On the other hand, the J1906+0335 results favor the latter model, with  $\ln\mathcal{B}=-3.38$. Per standard scales \citep{Kass+1995,Jeffreys1998}, the results for J1906+0335 and J1917+1142 are not notable, while the results for J1905+0413, J1929+1154 and J2010+3147 are ``decisive''. In short, the three sources with the most pulses clearly favor the log-normal distribution, while the other two are inconclusive.

\section{Spectra} \label{sec:spectra}

Pulsar radio spectra are usually well-approximated by power laws, $S(f)\propto f^{\alpha}$, with the spectral law index typically lying in the range $-2\leq\alpha\leq-1$ \citep{Bates+2013,Jankowski+2018}. If a RRAT is detected at two frequencies $f_1$ and $f_2$, its spectral index can be estimated through computing the mean single-pulse flux density at each frequency $\bar{S}_1\equiv\langle S(f_1)\rangle$ and $\bar{S}_2\equiv\langle S(f_2)\rangle$:
\begin{equation}
\alpha=\frac{\log_{10}(\bar{S}_1/\bar{S}_2)}{\log_{10}(f_1/f_2)}
\end{equation}
Since our CHIME detections yielded nondetections, we can only place lower limits on each RRAT's $\alpha$, by setting $\bar{S}_1$ to be the mean of its single-pulse flux densities at L-band\footnote{While L-band observations at Arecibo were centered at multiple frequencies because of the use of ALFA and the LBW, they are close enough that for an approximation they were at essentially the same frequency.} and replacing $\bar{S}_2$ with the sensitivity threshold in the CHIME band. The sources were too faint in the upper portions of the band to perform spectral analysis by subbanding the Arecibo data.

Computing the flux density of an individual pulse is done following the prescription described in \citealt{Miller2013}. We fit a Gaussian to the detrended time series, as described in Section~\ref{subsec:detrending}, and calculate a fitted signal-to-noise ratio by dividing the amplitude of the Gaussian by the standard deviation of the off-pulse region. This is then converted to a flux density by multiplying it by the expected radiometer noise:
\begin{equation}
\sigma_{\mathrm{rn}} = \frac{\beta\;\mathrm{SEFD}}{\sqrt{n_p\Delta ft_{\mathrm{samp}}}}
\end{equation}
where $\beta\simeq1$ is a factor describing the loss of sensitivity due to digitization.

Of the four sources observed with CHIME, J0623+15 yielded only one pulse in the original Arecibo data, insufficient to place meaningful constraints on its spectral index. The other three all yielded at least four pulses. Their mean L-band flux densities and spectral index limits are listed in Table~\ref{tab:spectra}. One of the 14 pulses from J0625+12 yielded a very poorly constrained amplitude and was therefore omitted.

Our power law index constraints are consistent with what we know about the pulsar population in general. Most pulsars have spectral indices in the interval $-2\leq\alpha\leq-1$ \citep{Bates+2013}, a range comfortably above the lower limits on $\alpha$ for each of the three pulsars. However, we refrain from drawing too firm conclusions, since these limits are derived from small numbers of pulses.

\begin{table}
  \renewcommand\thetable{7}
  \centering
  \setlength{\tabcolsep}{1mm}
  \begin{tabular}{c c c c}
  \hline
Source & $N_p$ & $S_1$ (mJy) & $\alpha_{\mathrm{min}}$\\
\hline\hline
J0625+12 & 13 & 9.6 & $-4.27$ \\
J1843+05 & 4 & 20 & $-3.03$ \\
J1924+10 & 11 & 39 & $-2.57$ \\
\hline
\end{tabular}
{\footnotesize
  \caption{The results of constraints on the spectral indices of three sources observed with both the Arecibo telescope and CHIME. $N_p$ is the number of pulses used, $S_1$ is the mean flux density of single pulses at L-band, and $\alpha_{\mathrm{min}}$ is the minimum possible power law index. The value of $S_1$ is below the nominal ALFA/Mock and LBW/PUPPI thresholds, but not by a large amount.}}
\end{table} \label{tab:spectra}

\section{Wait times} \label{sec:wait-times}

\subsection{A Poisson process?}

The single-pulse emission from several RRATs has been observed to be inconsistent with a Poisson process (see e.g. \citealt{McLaughlin+2009,Keane+2010,Keane+2011a,Palliyaguru+2011}. The pulse-to-pulse wait times from a Poisson process should follow an exponential distribution:
\begin{equation}
\mathcal{P}(\delta|r)=re^{-\delta r}
\end{equation}
with $\delta$ a wait time and $r$ the mean pulse rate. The deviations seen in some RRATs manifest as clusters of pulses. In the case of J1514-59, the distribution of wait times is bimodal, with one peak at times of a few pulse periods and one peak at times of several hundred pulse periods \citep{Keane+2010}. The short-interval portion of the distribution does appear roughly Poissonian. This motivates a study of the inter-pulse wait times from the RRATs with sufficient numbers of pulses, namely, J1905+0413, J1906+0335, J1917+1142, J1929+1154 and J2010+3147.

\subsection{Fitting wait time distributions}

Several different distributions have been fit to RRAT wait times. \citet{Shapiro-Albert+2018} noticed bumps in the wait time distribution of several RRATs at times on the order of tens of rotation periods, and so fit three additional distributions: exponential plus a Gaussian, an exponential plus a Maxwell-Boltzmann distribution, and a log-normal distribution plus a Maxwell-Boltzmann distribution. No such bumps are seen observed in our distributions. However, we allow for deviations from a Poisson process through the use of the Weibull distribution. The Weibull distribution generalizes the Poisson distribution to allow for clustering by introducing a shape parameter $k$, and has been applied to bursts from FRB 121102 \citep{Oppermann+2018}. For a rate $r$ and shape parameter $k$, the Weibull distribution is
\begin{equation}
\mathcal{W}(\delta|k,r)=k\delta^{-1}\left[\delta r\Gamma(1+1/k)\right]^ke^{-[\delta r\Gamma(1+1/k)]^k}
\end{equation}
where $\Gamma(x)$ is the gamma function. The likelihood function of an observation with length $\Delta$ with no pulses detected is \citep{Oppermann+2018}
\begin{equation}
\mathcal{L}_{\mathrm{obs}}(N=0|k,r)=\frac{\Gamma_i\left(1/k,\left[\Delta r\Gamma(1+1/k)\right]^k\right)}{k\Gamma(1+1/k)}
\end{equation}
where $\Gamma_i(x)$ is the incomplete gamma function. The likelihood function of an observation with $N$ pulses detected at times $t_1,t_2,...t_N$ is the product of $N+1$ likelihoods: the probability of not detecting any pulses before $t_1$, the probability of each of the successive wait times, and the probability of not detecting and pulses after $t_N$. If the observation again has length $\Delta$, this single-observation likelihood is
\begin{equation}
\begin{aligned}
\mathcal{L}_{\mathrm{obs}}(t_1,t_2,...,t_N|k,r)=\;&r\;\mathrm{CDF}(t_1|k,r)\\
&\times\prod_{i=1}^N\mathcal{W}(t_{i+1}-t_i|k,r)\\
&\times\;\mathrm{CDF}(\Delta-t_N|k,r)
\end{aligned}
\end{equation}
where
\begin{equation}
\mathrm{CDF}(\delta|k,r)=e^{-[\delta r\Gamma(1+1/k)]^k}
\end{equation}
which is $1$ minus the traditional definition of the cumulative distribution function (CDF).

The likelihood of the entire observing campaign is then the product of all of the single-observation likelihoods, assuming the observing cadence is much longer than the typical inter-pulse spacing, which is true for all of our sources with sufficient numbers of pulses to perform this analysis. We also follow the rest of the procedure outlined in \citealt{Oppermann+2018} to perform our fits, choosing log-uniform priors on $r$ and $k$. To enable comparisons between RRATs, we make $r$ dimensionless by dividing pulse times and observation lengths by the spin period of each source \citep{Shapiro-Albert+2018}. We perform MCMC fitting to obtain posteriors, as done in Section~\ref{sec:pulse-energies}.

A benefit of the Bayesian method to wait time fitting is that observations with non-detections are factored into the analysis. Typical works fitting wait time distributions (e.g. \citealt{Shapiro-Albert+2018}) in the wait times and fit curves to the resulting histograms. In addition to avoiding the systematic errors these induce, the Bayesian approach ensures that even non-detections contribute to the fits.

Table~\ref{tab:wait-times} shows the results of the fits. All RRATs yielded shape parameters in the range $-0.18\leq\ln k\leq-0.07$ corresponding to $0.84\leq k\leq0.93$. This indicates mild clustering; for comparison, \citealt{Oppermann+2018} measured $k=0.34^{+0.06}_{-0.05}$ for FRB 121102. However, J1906+0335 and J1917+1142 have posteriors consistent with $k=1$ (i.e. no clustering) within $2\sigma$, and no source has is dramatically inconsistent from $k=1$. We therefore conclude that there is at best mild evidence for clustering relative to a Poisson process.

\begin{table}
  \renewcommand\thetable{8}
  \centering
  \setlength{\tabcolsep}{1mm}
  \begin{tabular}{c c c}
  \hline
Source & $\ln(rP)$ & $\ln k$ \\
\hline\hline
J1905+0413 & $-4.90^{+0.10}_{-0.10}$ & $-0.18^{+0.06}_{-0.06}$ \\
J1906+0335 & $-4.38^{+0.13}_{-0.13}$ & $-0.13^{+0.09}_{-0.09}$ \\
J1917+1142 & $-5.28^{+0.14}_{-0.15}$ & $-0.10^{+0.10}_{-0.10}$ \\
J1929+1154 & $-2.82^{+0.06}_{-0.07}$ & $-0.11^{+0.04}_{-0.04}$ \\
J2010+3147 & $-2.65^{+0.04}_{-0.04}$ & $-0.07^{+0.03}_{-0.02}$ \\
\hline
\end{tabular}
{\footnotesize
  \caption{Results of fitting Weibull distributions to the wait time distributions of the five RRATs with significant numbers of pulses.}}
\end{table} \label{tab:wait-times}

\begin{table*}
  \renewcommand\thetable{9}
  \centering
  \setlength{\tabcolsep}{1mm}
  \begin{tabular}{c c c c c c c c c}
  \hline
Source & GPPS name & GPPS & PALFA DM & GPPS DM & PALFA $P$ & GPPS $P$ & Reference \\
& & label & (pc cm$^{-3}$) & (pc cm$^{-3}$) & (s) & (s) \\
\hline\hline
J0623+15 & J0623+1536 & Sparse & 92.5 & 92.7 & -- & 2.638545 & \citep{GPPS2} \\
J0625+12 & J0625+1254 & RRAT & 102 & 102.8 & -- & -- & \citep{GPPS2} \\
J1843+05 & J1843+0527 & Null & 262.3 &  261.1 & 2.035 & 2.034918 & \citep{GPPS2} \\
J1905+0413 & J1905+0414 & Sparse & 381 & 383.0 & 0.894 & 0.894124 & \citep{GPPS2} \\
J1906+0335 & J1906+0335g & Sparse & 212 & 210.5 & 1.296 & 1.29639 & \citep{GPPS2,GPPS6} \\
J1917+1142 & J1916+1142B & RRAT & 319 & 317.7 & 1.188 & 1.18795 & \citep{GPPS2} \\
J1924+10 & J1924+1006 & Sparse & 176.8 & 178.1 & 4.620 & 4.619757 & \citep{GPPS2} \\
J1929+1154 & J1929+1155 & Sparse & 80 & 81.2 & 3.218 & 3.216892 & \citep{GPPS2} \\
J2010+3147 & J2010+31 & Pulsar & 251 & 251.8 & 1.551 & 1.551535 & \citep{GPPS1} \\
\hline
\end{tabular}
{\footnotesize
  \caption{A list of the PALFA single-pulse sources in this sample that may have been redetected by GPPS. The classification abbreviations here refer to categories of pulsars established by GPPS; here, we use ``Sparse" to refer to a ``weak pulsar with strong single pulses" \citep{GPPS2} and ``Null" to refer to a ``pulsar with nulling features in FAST observations" \citep{GPPS2}. Note that J1906+0335g and J1916+1142B were inadvertently believed to be newly-discovered pulsars, but correspond to the previously-known PALFA sources J1906+0335 and J1917+1142, respectively.}}
\end{table*} \label{tab:gpps}

\subsection{Longer-timescale variations}

In addition to the variations in observed pulse rate expected from a source well-described by a Poisson (or Weibull) process, some of the RRATs appear to show clustering. J1859+07, for example, yielded two detections in its first four follow-up observations, which fell within a 90-day span, but no pulses in the remaining 23 observations, which fell within a 1122-day span. Similarly, J1929+1154 exhibited an excess of pulses in four observations between MJD 56586 and MJD 56590 (94 in 60 minutes) and six observations between MJD 56936 and MJD 56950 (74 in 30 minutes) and only 130 pulses in the remaining 23 observations (195 minutes). These periods of apparent increased and decreased activity, if not simple due to random variations, could be due to an intrinsic process or to propagation effects in the interstellar medium, such as refractive scintillation or extreme scattering events.

\begin{figure*}
    \centering
    \includegraphics[width=1.9\columnwidth]{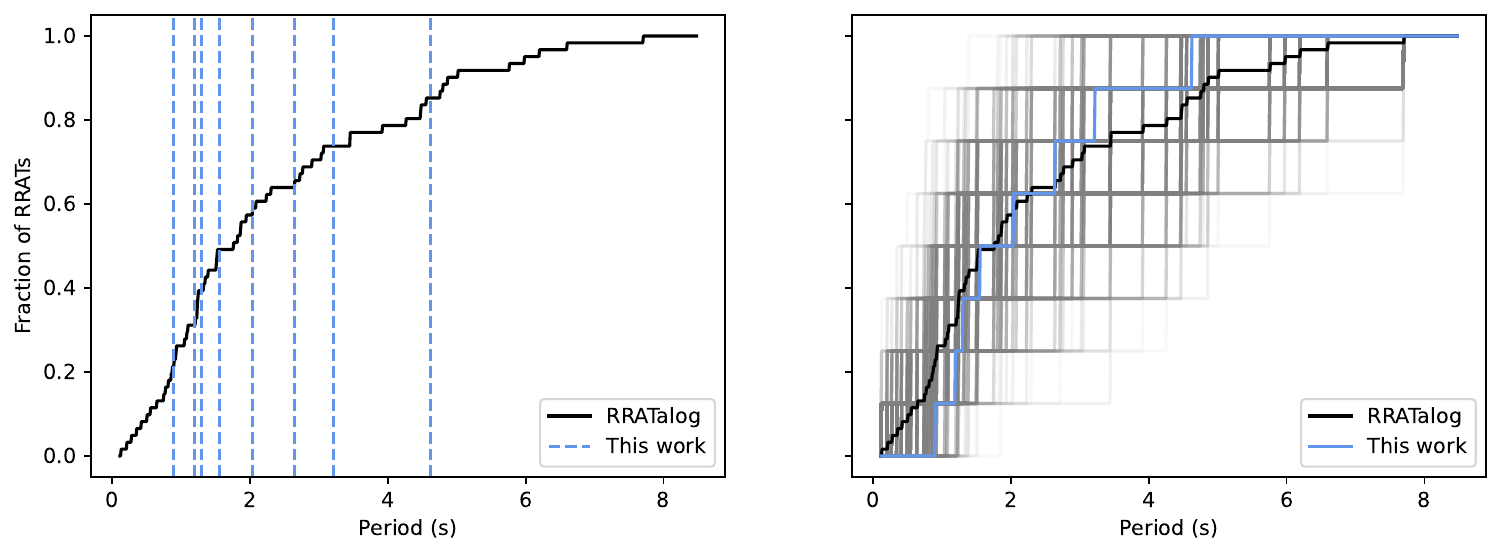}
    \caption{The left panel shows the cumulative distribution of periods of sources in the RRATalog (black) compared with the periods of the sources in our sample (blue). The right panel shows the RRATalog CDF (black), the sample CDF (blue) and the CDFs of each of the $1000$ simulations (grey).}
    \label{fig:period-distribution}
\end{figure*}

\section{Discussion} \label{sec:discussion}

\subsection{Population statistics} \label{subsec:population}

The properties of the RRATs in this sample appear consistent with those of the the RRAT population as a whole. Figure~\ref{fig:period-distribution} shows the cumulative distribution function of RRAT periods, taken from the RRATalog (Agarwal et al. in prep.), along with the cumulative distribution function of the eight periods known from this sample.\footnote{For J0623+15, we again use the period from the (GPPS; see Section~\ref{subsec:GPPS}).} The Kolmogorov-Smirnov (KS) statistic of the sample is $\mathrm{KS}=0.213$. We performed $1000$ bootstrap simulations of 8 draws from the RRATalog period distribution, computing a CDF and KS statistic for each simulation. The sample's KS statistic is lower than that of $792$ of the simulations, implying a $p$-value of $p\approx0.8$. This is clearly not significant, indicating that we observe no significant deviations from the observed RRAT period distribution.

\begin{figure*}
\centering
\includegraphics[width=1.9\columnwidth]{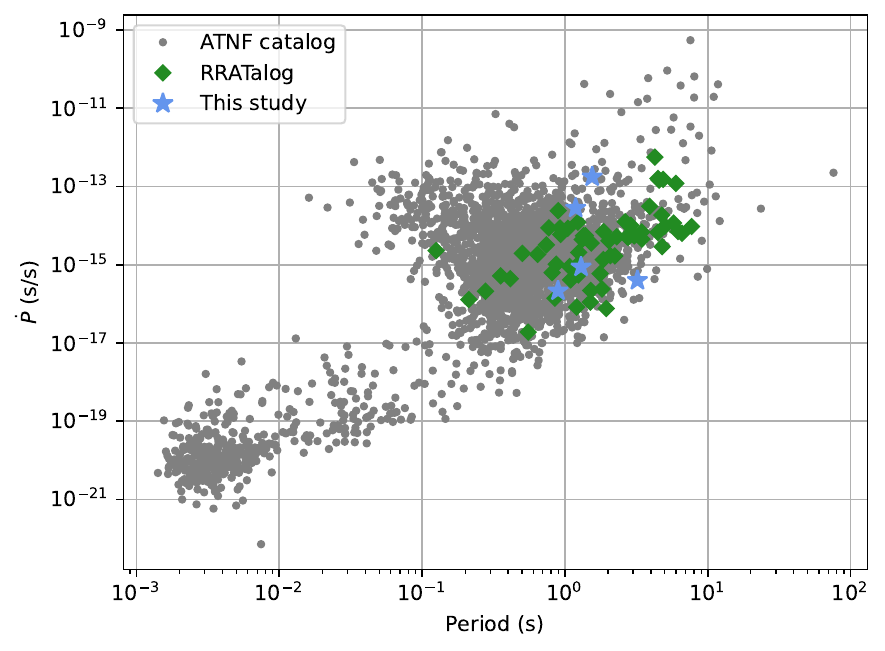} \label{fig:p-pdot-diagram}
\caption{A $P$--$\dot{P}$ diagram featuring pulsars from the ATNF catalog, rotating radio transients from the RRATalog, and five sources from this study (J1905+0413, J1906+0335, J1917+1142, J1929+1154 and J2010+3147). The ATNF data is from version 2.6.5 of the catalog.}
\end{figure*}

Figure~\ref{fig:p-pdot-diagram} shows a $P$--$\dot{P}$ diagram, featuring periods and period derivatives taken from the Australia Telescope National Facility (ATNF) pulsar catalog \citep{ATNFcatalog}\footnote{\url{https://www.atnf.csiro.au/research/pulsar/psrcat/}}, the RRATalog and the four sources for which we obtained timing solutions. J1905+0413, J1906+0335 and J1929+1154 appear to fall well within the existing RRAT population, while J2010+3147 features a rather high $\dot{P}$. However, the former three sources do not have well-constrained period derivatives, making any further conclusions tentative at best.

\subsection{Comparison with GPPS} \label{subsec:GPPS}

As noted before, the Galactic Plane Pulsar Snapshot survey appears to have redetected a sizable number of our single-pulse sources. Table~\ref{tab:gpps} lists the nine PALFA pulsars from this work with likely GPPS counterparts. A possible counterpart based on dispersion measure and position was also identified for J1859+07 (J1859+0658g; \citealt{GPPS1,GPPS8}). However, we believe these to be different objects, as they differ in dispersion measure by a substantial amount ($\Delta\mathrm{DM}=12.6$ pc cm$^{-3}$). We therefore believe any association to be unlikely.


\subsubsection{RRATs or pulsars?} \label{subsubsec:rrats-or-pulsars}

All sources discussed here were initially detected in single-pulse searches of PALFA data. J0613+18 is in fact likely a fast radio burst (see Section~\ref{subsec:J0613+18-nature}) and J2010+3147 was regularly detected in periodicity searches in follow-up observations. The remaining ten sources are primarily (and, with the exception of J1929+1154, only) detected in PALFA data through single-pulse searches, and can therefore be classified as ``PALFA RRATs".

Of those ten, six (J0623+15, J1843+05, J1905+0413, J1906+0335, J1924+10 and J1929+1154) were detected by GPPS in periodicity searches and were classified by the GPPS team as either ``weak pulsar[s] with strong single pulses" \citep{GPPS2} or nulling pulsars. The remaining four (J0529+25, J0625+12, J1859+07 and J1917+1142) have only been detected by any survey through single pulses. They may also be responsible for continuous emission, but we cannot rule out the possibility that they are truly, intrinsically, single-pulse sources.

\subsection{J0613+18: Galactic or extragalactic?} \label{subsec:J0613+18-nature}

The dispersion measure of J0613+18 is $\sim$~400 pc cm$^{-3}$, well exceeding the maximum dispersion measures predicted by the NE2025 and YMW16 models from the Galactic disk along its line of sight (159.1 pc cm$^{-3}$ and 297 pc cm$^{-3}$, respectively). Such a DM could be consistent with a location within the Milky Way if a structure not included in the electron density models -- potentially a nebula or HII region -- lies along the line of sight. A previously-conducted multiwavelength search  yielded no such structures, leading to the conclusion that the source is likely extragalactic \citep{Patel+2018}. The newer data presented here do not substantially change the conclusion that J0613+18 likely lies outside the Milky Way; for completeness, however, we discuss two possible sources of error in making a determination of Galactic or extragalactic origin.

First, the Galactic halo often has a significant contribution to the DMs of objects far beyond the Galactic disk. Recent studies have found median halo contributions to fast radio burst dispersion measures of $\sim40$--60 pc cm$^{-3}$, with some FRBs showing even higher contributions \citep{Yamasaki+Tomonori2020,Das+2021}. It is not inconceivable that a pulsar or RRAT located deep in the halo could have a dispersion measure perhaps even $\sim100$ pc cm$^{-3}$ higher than the limit predicted by the Galactic electron density models. However, along this line of sight, the halo is expected to contribute only $\sim41$ pc cm$^{-3}$ according to the YT20 halo model \citep{Yamasaki+Tomonori2020} as implemented in the \texttt{PyGEDM} package \citep{Price+2021}.

It is also possible that the Galactic electron density models perform poorly in this region of the sky, and underestimate the contribution to dispersion measure from the Milky Way's disk towards J0613+18. In fact, however, the opposite is true: YMW16 is known to \textit{overestimate} the disk DM contribution towards the Galactic anticenter, close to the sky position of J0613+18 \citep{Price+2021}. This explains why the maximum YMW16 DM is significantly higher than the maximum NE2025 DM. Furthermore, no pulsar listed in the ATNF catalog within ten degrees of J0613+18 has a DM of even 200 pc cm$^{-3}$. Without an intervening structure, significant halo contribution, or known model bias, we find no convincing evidence for J0613+18 lying within the Milky Way.

\citealt{Patel+2018} placed an upper limit on the source's redshift of $z\lesssim0.15$, a distance of $\lesssim0.6$ Gpc. We searched for a possible host in the NASA/IPAC Extragalactic Database (NED). The half power beamwidth of the ALFA receiver was 3.6$^\prime$; to include sidelobes, we searched out to 5$^\prime$. The database returned 659 sources, including J0613+18. Of the remaining 658, 657 are designated as ``infrared sources" and one, WISEA J061249.18+184618.2, is identified as a galaxy.\footnote{It is also possible that one or more of the ``infrared sources" are in fact galaxies, and could be associated with J0613+18.} WISEA J061249.18+184618.2's redshift is unknown, and it is separated from J0613+18 by approximately 2.7$^\prime$. An association between the two is plausible but currently not convincing based on the available data.

\subsection{Future work} \label{subsec:future-work}

As noted above, the loss of the 305-meter telescope at Arecibo prematurely ended the PALFA survey. As with other projects at Arecibo, this has severely impeded our follow-up campaigns. Some of these sources are only known from faint, low-SNR pulses, and should be visible only from a select few telescopes; as such, obtaining telescope time to perform follow-up observations is understandably difficult. However, the recent GPPS results showing that many of the PALFA RRAT candidates actually emit periodic emission imply that the detection prospects from campaigns at highly sensitive telescopes are better than were previously apparent. Possible telescopes for follow-up observations include FAST and the Green Bank Telescope (GBT), as well as future next-generation facilities like the Deep Synoptic Array 2000 (DSA-2000) and Square Kilometer Array (SKA). It is also possible that some of these sources may be serendipitously redetected in blind surveys by one of these telescopes, which would further motivate follow-up campaigns at the facility.

\subsubsection{Statistics}

In a single-pulse study, more pulses generally improve the quality of results. For example, in the case of a source emitting with Poisson statistics has a pulse rate with fractional uncertainty $\sigma_{\mathcal{R}}/\mathcal{R}=(\sqrt{N}/T)/(N/T)=1/\sqrt{N}$. More observations of a source should therefore increase the accuracy and precision of pulse rate estimates. In our data set, J1905+0413 and J1924+10 have almost exactly the same pulse rate (0.441 pulses per minute), yet the uncertainty in the pulse rate of J1905+0413 is lower by a factor of approximately four because over an order of magnitude more pulses were detected from it. The same reasoning holds for other estimated parameters, such as the best-fit values of the energy and wait time distributions. 

\subsubsection{Timing} \label{subsubsec:timing}

Further follow-up observations at Arecibo would have significantly improved the timing results. A longer timing baseline would have lowered the root mean square of the residuals and enabled the measurement of astrometric parameters and higher spin derivatives. Additionally, more observations would have meant more pulses, reducing the noise contaminating each template and therefore the TOA uncertainties. Finally, while J1843+05 and J1924+10 were not the most active sources in this sample, they were active enough that they could likely have been timed with follow-up campaigns of reasonable duration.

Timing observations at other telescopes would also be helpful. A facility capable of detecting any of these sources in a periodicity search could produce TOAs from folded profiles, which would complement and potentially improve on single-pulse timing.

\subsubsection{High-priority targets}

J1843+05 and J1924+10 should arguably be the highest-priority targets for any future timing campaigns. They are not particularly sporadic, with pulse rates among the highest of this sample. The only problem prohibiting timing was the small number of observations. A hypothetical follow-up campaign at Arecibo of the same kind as those for J1905+0413, J1906+0335, J1917+1142 and J1929+1154 would presumably be sufficient. Follow-up observations would also enable a study of the nulling observed in J1843+05 by \citealt{GPPS2}.

At the other end of the spectrum, the extremely intermittent sources are also of interest. J0529+25, J0625+12 and J1859+07 are all quite sporadic and, like J1917+1142, have not yet been detected in periodicity searches. While it is impossible to rule out the presence of periodic emission, it is possible to put lower and lower limits on the mean flux density of pulsed, periodic emission from a source. It would be beneficial to see whether more sensitive surveys can detect any of these objects in periodicity searches. Multifrequency observations could also provide support for models producing transient emission. For example, gamma-ray emission from an RRAT undergoing propeller spindown might be visible \citep{Li2006}. Circumpulsar asteroids from a debris disk could be detected via infrared emission from the disk or via modulation of high-energy emission \citep{Cordes+2008}. 

J0613+18, which we have classified as a fast radio burst, is also a target of interest, albeit a highly risky one. While it quite likely lies outside the Milky Way, we have not been able to rule out a Galactic origin. Further observations might aid in this. Unfortunately, as of 2025, only $\sim8$\% of FRBs have been detected as repeaters\footnote{72 out of 889, per the Blinkverse database (\url{https://blinkverse.zero2x.org/}; \citealt{Blinkverse}), as of 16 September 2025.} \citep{Blinkverse}, and, given its burst morphology, J0613+18 is likely not among them, as discussed in Section~\ref{subsec:J0613+18-profile}. This means that follow-up observations targeting it would be a high-risk, high-reward undertaking.

\section{Conclusion} \label{sec:conclusion}

We present here an analysis of twelve more sources discovered in single-pulse searches of observations from the PALFA survey at the Arecibo Observatory. Upon further examination, one appears to be a fast radio burst and one appears to be a normal pulsar. After comparison with independent observations by the Galactic Plane Pulsar Snapshot survey, six appear to be pulsars with either strong single pulses that are part of a continuous intensity distribution or significant nulling. The remaining four might still, potentially, show intermittent single pulses isolated in time.

For the five sources with sufficient numbers of pulses, we obtained phase-connected timing solutions using TOAs from single pulses. For those five and two others, we also studied the distributions of pulse energies and/or inter-pulse wait times, showing similarities with previous studies of other pulsars and RRATs.

The results also underline the importance of single-pulse searches in pulsar surveys. Standard periodicity searches alone would have missed many of the objects in this sample have been found by other surveys to display periodic emission. Any survey performing periodicity searches should also perform single-pulse searches, which could find $\sim15$\% of all survey candidates \citep{GPPS2}.

\subsection*{Software}

This work has made use of several software packages, including \texttt{APTB} \citep{APTB}, \texttt{Astropy} \citep{astropy1,astropy2,astropy3}, \texttt{DRACULA} \citep{DRACULA}, \texttt{emcee} \citep{emcee}, \texttt{harmonic} \citep{harmonic}, \texttt{NumPy}, \citep{NumPy}, \texttt{PINT} \citep{PINT}, \presto\;\citep{Ransom2011}, \texttt{SciPy} \citep{SciPy}, and \tempo \citep{Tempo2}.

\acknowledgments

\section*{Acknowledgements}

GMD thanks Jackson Taylor for valuable discussions about using APTB. GMD was supported through NSF award \# 2406570.

This research has made use of the NASA/IPAC Extragalactic Database (NED), which is funded by the National Aeronautics and Space Administration and operated by the California Institute of Technology, and NASA's Astrophysics Data System. This work also has made use of Astropy\footnote{\url{http://www.astropy.org}}, a community-developed core Python package and an ecosystem of tools and resources for astronomy \citep{astropy1,astropy2,astropy3}.

\section*{Data Availability}

Raw PSRFITS files and code are available from the authors upon reasonable request, as well as intermediate data products, including diagnostic plots and single pulse profiles. Example code and some single pulse profiles are available on GitHub at \url{https://github.com/GrahamDoskoch/PALFA-RRAT-example-code}.

\bibliography{rrat.bib}{}
\bibliographystyle{aasjournal}

\appendix

\section{Energy fitting details} \label{sec:energy-fitting-appendix}

In our comparison, we are interested in two things: accuracy and precision. Inaccuracy could manifest as a systematic offset in parameter estimation in one direction or another, such as consistently overestimating the value of some parameter $\theta_i$. Imprecision could manifest as a broader spread in the estimates of $\theta_i$; a wider distribution should reduce confidence in the method's efficacy.

For each of the two distributions of interest, we perform 100 simulations, drawing $N=51$ pulses from the distribution, using parameters typical of the sources in our sample ($\mu=0$ and $\ln\sigma=-0.50$ for the log-normal distribution, and $b=2.00$ and $\ln c=1.00$ for the power law/exponential distribution). We then apply both of the fitting methods to each realization and create distributions of the estimates of the parameters.

\begin{figure}[h]
\centering
\begin{minipage}{0.8\columnwidth}
  \centering
  \includegraphics[width=0.8\columnwidth]{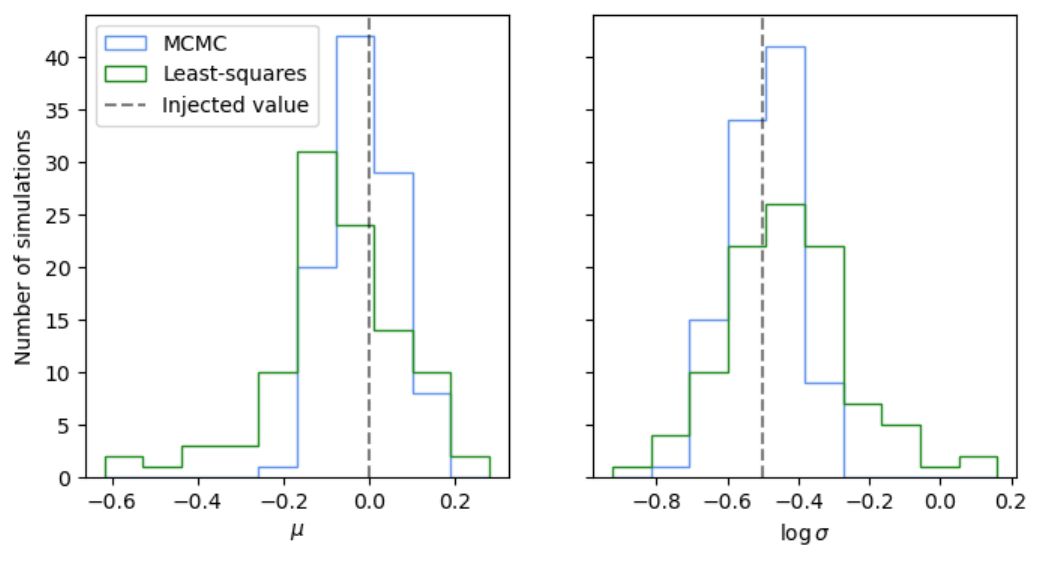}
  \label{fig:mu-sig-comparison-plot}
\end{minipage}%
\begin{minipage}{0.8\columnwidth}
  \centering
  \includegraphics[width=0.8\columnwidth]{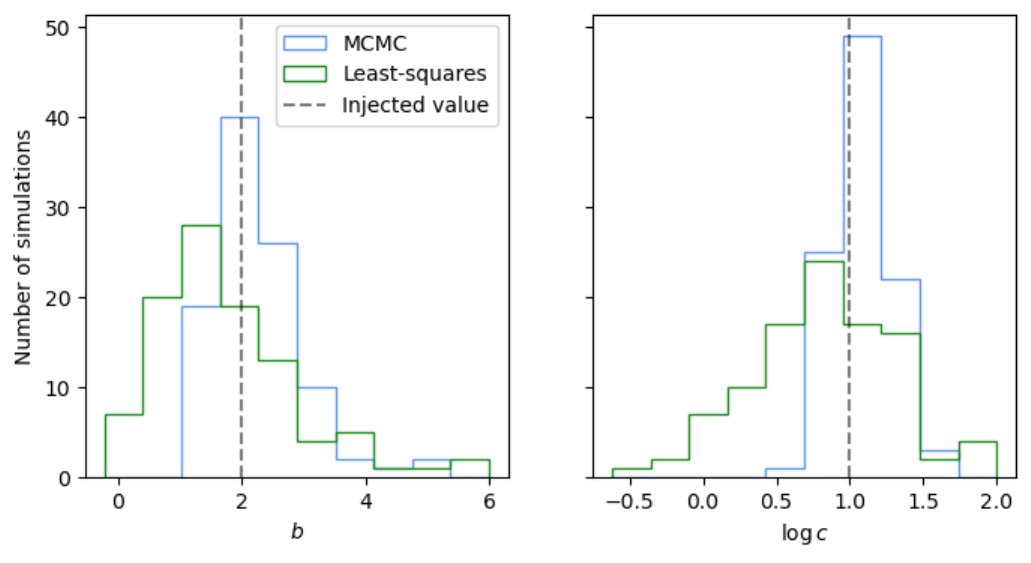}
  \label{fig:b-c-comparison-plot}
\end{minipage}
\caption{The results of the two energy fitting methods applied to 100 simulations of $N=51$ pulses drawn from log-normal (top) and power law/exponential (bottom) distributions. Largely, the MCMC method outperforms the least-squares fitting in both accuracy and precision.} \label{fig:comparison-simlations}
\end{figure}

The results are shown in Figure~\ref{fig:comparison-simlations}. For all four parameters, we find that the distribution of fits are narrower for the MCMC fits by a factor of $\sim2$ and are clearly centered around the injected values, implying the MCMC method is notably more accurate and more precise than the traditional least squares approach. This strongly reinforces the notion that, at least for small sample sizes, the MCMC approach should be used instead of binning and least squares fitting.

\newpage

\section{Observation details} \label{sec:obs-details-appendix}

Table~\ref{tab:observation-details} describes the results of the single-pulse searches of each source. Additional information is available upon request.

\begin{longtable*}{c|c|c|c}
\caption{Details of each observation from the campaign at Arecibo. Observations in Table~\ref{tab:observation-details} using the ALFA/Mock combinations are marked with a $^{\dagger}$; observations using the WAPP/Mock combination are marked with a $^{\ddagger}$.} \label{tab:observation-details}\\\hline
Source & Date & Length & Number of pulses \\
& (MJD) & (minutes) & \\\hline\hline
\endhead
J0529+25 & 58071 & 15.00 & 0 \\
J0529+25 & 58216 & 13.25 & 0 \\
J0529+25 & 58218 & 15.00 & 1 \\
J0529+25 & 58220 & 15.00 & 0 \\
J0529+25 & 58677 & 15.00 & 0 \\
J0529+25 & 58706 & 15.00 & 0 \\
J0529+25 & 58735 & 15.00 & 0 \\
J0529+25 & 58737 & 15.00 & 0 \\
J0529+25 & 58739 & 15.00 & 0 \\
J0529+25 & 58749 & 15.00 & 0 \\
J0529+25 & 58754 & 15.00 & 0 \\
J0529+25 & 58764 & 15.00 & 0 \\
J0529+25 & 58828 & 15.00 & 0 \\
J0529+25 & 58897 & 15.00 & 0 \\
J0529+25 & 58922 & 20.00 & 0 \\
J0529+25 & 58953 & 15.00 & 0 \\
J0529+25 & 59013 & 15.00 & 0 \\
J0529+25 & 59042 & 15.00 & 0 \\
J0529+25 & 59070 & 17.00 & 0 \\
\hline
J0529+25 & Total & 290.27 & 1 \\
\hline
J0613+18 & 56974$^{\dagger}$ & 3.27 & 1 \\
J0613+18 & 57984 & 23.33 & 0 \\
J0613+18 & 58071 & 18.33 & 0 \\
J0613+18 & 58101 & 23.33 & 0 \\
J0613+18 & 58153 & 15.00 & 0 \\
J0613+18 & 58157 & 3.28 & 0 \\
J0613+18 & 58177 & 15.00 & 0 \\
J0613+18 & 58189 & 3.28 & 0 \\
J0613+18 & 58191 & 20.00 & 0 \\
J0613+18 & 58233 & 20.00 & 0 \\
J0613+18 & 58243 & 21.61 & 0 \\
J0613+18 & 58297 & 20.00 & 0 \\
J0613+18 & 58300 & 20.00 & 0 \\
J0613+18 & 58300 & 60.00 & 0 \\
J0613+18 & 58736 & 20.00 & 0 \\
J0613+18 & 58737 & 15.00 & 0 \\
J0613+18 & 58739 & 15.00 & 0 \\
J0613+18 & 58741 & 20.00 & 0 \\
J0613+18 & 58744 & 16.75 & 0 \\
J0613+18 & 58749 & 15.00 & 0 \\
J0613+18 & 58754 & 15.00 & 0 \\
J0613+18 & 58764 & 15.00 & 0 \\
J0613+18 & 58795 & 20.00 & 0 \\
J0613+18 & 58828 & 15.00 & 0 \\
J0613+18 & 58897 & 15.00 & 0 \\
J0613+18 & 58953 & 15.00 & 0 \\
J0613+18 & 58981 & 8.92 & 0 \\
J0613+18 & 59013 & 15.00 & 0 \\
J0613+18 & 59042 & 15.00 & 0 \\
\hline
J0613+18 & Total & 502.15 & 1 \\
\hline
J0623+15 & 57294$^{\dagger}$ & 3.28 & 1 \\
J0623+15 & 57354$^{\dagger}$ & 3.28 & 0 \\
J0623+15 & 57984 & 15.33 & 0 \\
J0623+15 & 58071 & 13.34 & 0 \\
J0623+15 & 58101 & 15.00 & 0 \\
J0623+15 & 58178 & 11.67 & 0 \\
J0623+15 & 58191 & 20.00 & 0 \\
J0623+15 & 58214 & 15.00 & 0 \\
J0623+15 & 58215 & 15.77 & 0 \\
J0623+15 & 58216 & 15.00 & 0 \\
J0623+15 & 58218 & 15.00 & 0 \\
J0623+15 & 58220 & 15.00 & 0 \\
J0623+15 & 58228 & 15.00 & 0 \\
J0623+15 & 58233 & 16.67 & 0 \\
J0623+15 & 58243 & 17.84 & 0 \\
J0623+15 & 58297 & 8.40 & 0 \\
J0623+15 & 58300 & 10.77 & 0 \\
J0623+15 & 58526 & 3.28 & 0 \\
J0623+15 & 58527 & 3.30 & 0 \\
J0623+15 & 58677 & 15.00 & 0 \\
J0623+15 & 58706 & 15.00 & 0 \\
J0623+15 & 58735 & 15.00 & 0 \\
J0623+15 & 58736 & 15.00 & 0 \\
J0623+15 & 58737 & 2.65 & 0 \\
J0623+15 & 58741 & 15.00 & 0 \\
J0623+15 & 58744 & 15.00 & 0 \\
J0623+15 & 58749 & 20.00 & 0 \\
J0623+15 & 58795 & 15.00 & 0 \\
J0623+15 & 58828 & 15.00 & 0 \\
J0623+15 & 58897 & 15.00 & 0 \\
J0623+15 & 58922 & 15.00 & 0 \\
J0623+15 & 58923 & 10.39 & 0 \\
J0623+15 & 58953 & 15.00 & 0 \\
J0623+15 & 58981 & 15.00 & 0 \\
J0623+15 & 59042 & 12.14 & 0 \\
\hline
J0623+15 & Total & 458.16 & 1 \\
\hline
J0625+12 & 57733$^{\dagger}$ & 3.27 & 1 \\
J0625+12 & 57984 & 15.00 & 1 \\
J0625+12 & 58071 & 5.63 & 1 \\
J0625+12 & 58101 & 15.00 & 0 \\
J0625+12 & 58153 & 13.69 & 0 \\
J0625+12 & 58178 & 11.67 & 0 \\
J0625+12 & 58191 & 6.24 & 0 \\
J0625+12 & 58215 & 6.62 & 0 \\
J0625+12 & 58243 & 8.69 & 0 \\
J0625+12 & 58297 & 16.67 & 0 \\
J0625+12 & 58300 & 16.67 & 0 \\
J0625+12 & 58677 & 15.00 & 0 \\
J0625+12 & 58706 & 15.00 & 0 \\
J0625+12 & 58735 & 25.45 & 0 \\
J0625+12 & 58736 & 15.00 & 0 \\
J0625+12 & 58739 & 14.20 & 0 \\
J0625+12 & 58741 & 15.00 & 0 \\
J0625+12 & 58744 & 15.00 & 0 \\
J0625+12 & 58754 & 20.96 & 0 \\
J0625+12 & 58764 & 20.00 & 3 \\
J0625+12 & 58795 & 15.00 & 1 \\
J0625+12 & 58811$^{\dagger}$ & 3.28 & 0 \\
J0625+12 & 58814$^{\dagger}$ & 3.28 & 0 \\
J0625+12 & 58828 & 15.00 & 2 \\
J0625+12 & 58922 & 15.00 & 1 \\
J0625+12 & 58923 & 18.00 & 1 \\
J0625+12 & 58981 & 15.00 & 1 \\
J0625+12 & 59013 & 14.40 & 1 \\
J0625+12 & 59042 & 18.00 & 1 \\
J0625+12 & 59070 & 10.23 & 0 \\
\hline
J0625+12 & Total & 401.96 & 14 \\
\hline
J1843+05 & 58295$^{\dagger}$ & 4.60 & 4 \\
J1843+05 & 59023 & 8.34 & 0 \\
J1843+05 & 59044 & 12.00 & 0 \\
\hline
J1843+05 & Total & 24.94 & 4 \\
\hline
J1859+07 & 57309$^{\dagger}$ & 4.60 & 1 \\
J1859+07 & 57505 & 14.50 & 1 \\
J1859+07 & 57539 & 14.50 & 0 \\
J1859+07 & 57558 & 13.50 & 0 \\
J1859+07 & 57595 & 15.00 & 1 \\
J1859+07 & 57624 & 15.00 & 0 \\
J1859+07 & 57653 & 15.00 & 0 \\
J1859+07 & 57670 & 10.00 & 0 \\
J1859+07 & 57671 & 10.00 & 0 \\
J1859+07 & 57689 & 5.00 & 0 \\
J1859+07 & 57699 & 15.00 & 0 \\
J1859+07 & 57716 & 15.00 & 0 \\
J1859+07 & 57748 & 15.00 & 0 \\
J1859+07 & 57774 & 10.00 & 0 \\
J1859+07 & 57802 & 10.73 & 0 \\
J1859+07 & 57838 & 10.00 & 0 \\
J1859+07 & 57864 & 15.00 & 0 \\
J1859+07 & 57892 & 10.00 & 0 \\
J1859+07 & 57924 & 10.00 & 0 \\
J1859+07 & 58680 & 15.00 & 0 \\
J1859+07 & 58711 & 10.00 & 0 \\
J1859+07 & 58727 & 10.00 & 0 \\
J1859+07 & 58728 & 10.00 & 0 \\
J1859+07 & 58731 & 10.00 & 0 \\
J1859+07 & 58733 & 10.00 & 0 \\
J1859+07 & 58736 & 10.00 & 0 \\
J1859+07 & 58741 & 18.37 & 0 \\
J1859+07 & 58745 & 10.00 & 0 \\
\hline
J1859+07 & Total & 331.24 & 3 \\
\hline
J1905+0413 & 57274$^{\dagger}$ & 4.60 & 6 \\
J1905+0413 & 57505 & 9.60 & 8 \\
J1905+0413 & 57595 & 15.00 & 8 \\
J1905+0413 & 57624 & 15.00 & 4 \\
J1905+0413 & 57653 & 15.00 & 6 \\
J1905+0413 & 57670 & 10.00 & 0 \\
J1905+0413 & 57671 & 10.00 & 5 \\
J1905+0413 & 57672 & 15.00 & 2 \\
J1905+0413 & 57674 & 10.00 & 1 \\
J1905+0413 & 57679 & 10.00 & 1 \\
J1905+0413 & 57699 & 15.00 & 0 \\
J1905+0413 & 57716 & 15.00 & 8 \\
J1905+0413 & 57748 & 15.00 & 10 \\
J1905+0413 & 57774 & 10.00 & 3 \\
J1905+0413 & 57838 & 10.00 & 6 \\
J1905+0413 & 57864 & 15.00 & 2 \\
J1905+0413 & 57892 & 10.00 & 3 \\
J1905+0413 & 57924 & 10.00 & 6 \\
J1905+0413 & 58007 & 10.00 & 0 \\
J1905+0413 & 58680 & 15.00 & 8 \\
J1905+0413 & 58711 & 10.00 & 3 \\
J1905+0413 & 58727 & 10.00 & 10 \\
J1905+0413 & 58728 & 10.00 & 10 \\
J1905+0413 & 58731 & 10.00 & 10 \\
J1905+0413 & 58733 & 10.00 & 3 \\
J1905+0413 & 58736 & 10.00 & 9 \\
J1905+0413 & 58740 & 10.36 & 0 \\
J1905+0413 & 58745 & 10.00 & 9 \\
\hline
J1905+0413 & Total & 319.61 & 141 \\
\hline
J1906+0335 & 56742 & 9.00 & 4 \\
J1906+0335 & 56771 & 10.00 & 0 \\
J1906+0335 & 56800 & 10.00 & 7 \\
J1906+0335 & 56801 & 10.00 & 7 \\
J1906+0335 & 56803 & 10.00 & 5 \\
J1906+0335 & 56805 & 10.00 & 7 \\
J1906+0335 & 56808 & 10.00 & 4 \\
J1906+0335 & 56813 & 10.00 & 6 \\
J1906+0335 & 56818 & 10.00 & 5 \\
J1906+0335 & 56828 & 10.00 & 8 \\
J1906+0335 & 56857 & 4.31 & 0 \\
J1906+0335 & 56947 & 10.00 & 7 \\
J1906+0335 & 56976 & 6.24 & 6 \\
J1906+0335 & 57001 & 4.69 & 4 \\
J1906+0335 & 57520 & 14.50 & 2 \\
J1906+0335 & 57556 & 14.50 & 0 \\
\hline
J1906+0335 & Total & 153.27 & 72 \\
\hline
J1917+1142 & 55438 & 4.62 & 0 \\
J1917+1142 & 55513$^{\dagger}$ & 4.62 & 3 \\
J1917+1142 & 56527 & 12.00 & 1 \\
J1917+1142 & 56555 & 15.00 & 4 \\
J1917+1142 & 56586 & 15.00 & 3 \\
J1917+1142 & 56587 & 15.00 & 0 \\
J1917+1142 & 56588 & 15.00 & 0 \\
J1917+1142 & 56590 & 15.00 & 3 \\
J1917+1142 & 56595 & 10.01 & 4 \\
J1917+1142 & 56599 & 15.00 & 2 \\
J1917+1142 & 56605 & 15.00 & 1 \\
J1917+1142 & 56617 & 15.00 & 1 \\
J1917+1142 & 56646 & 15.00 & 0 \\
J1917+1142 & 56746 & 15.00 & 6 \\
J1917+1142 & 56776 & 15.00 & 3 \\
J1917+1142 & 56811 & 15.00 & 0 \\
J1917+1142 & 56834 & 15.00 & 4 \\
J1917+1142 & 56895 & 15.00 & 0 \\
J1917+1142 & 56946 & 16.00 & 4 \\
J1917+1142 & 56973 & 16.00 & 6 \\
J1917+1142 & 57003 & 16.00 & 9 \\
\hline
J1917+1142 & Total & 289.27 & 54 \\
\hline
J1924+10 & 58370$^{\dagger}$ & 4.60 & 2 \\
J1924+10 & 59023 & 8.34 & 3 \\
J1924+10 & 59044 & 12.00 & 6 \\
\hline
J1924+10 & Total & 24.94 & 11 \\
\hline
J1929+1154 & 54907$^{\dagger}$ & 4.65 & 4 \\
J1929+1154 & 56189 & 10.00 & 1 \\
J1929+1154 & 56477 & 5.00 & 3 \\
J1929+1154 & 56527 & 12.00 & 11 \\
J1929+1154 & 56556 & 15.00 & 8 \\
J1929+1154 & 56586 & 15.00 & 18 \\
J1929+1154 & 56587 & 15.00 & 33 \\
J1929+1154 & 56588 & 15.00 & 15 \\
J1929+1154 & 56590 & 15.00 & 28 \\
J1929+1154 & 56595 & 10.00 & 4 \\
J1929+1154 & 56599 & 15.00 & 19 \\
J1929+1154 & 56605 & 15.00 & 14 \\
J1929+1154 & 56617 & 10.00 & 8 \\
J1929+1154 & 56646 & 10.00 & 13 \\
J1929+1154 & 56746 & 10.00 & 7 \\
J1929+1154 & 56776 & 10.00 & 5 \\
J1929+1154 & 56811 & 10.00 & 5 \\
J1929+1154 & 56834 & 10.00 & 0 \\
J1929+1154 & 56854 & 6.00 & 7 \\
J1929+1154 & 56856 & 6.00 & 3 \\
J1929+1154 & 56879 & 6.00 & 6 \\
J1929+1154 & 56935 & 5.00 & 0 \\
J1929+1154 & 56936 & 5.00 & 10 \\
J1929+1154 & 56937 & 5.00 & 7 \\
J1929+1154 & 56939 & 5.00 & 11 \\
J1929+1154 & 56941 & 5.00 & 15 \\
J1929+1154 & 56945 & 5.00 & 10 \\
J1929+1154 & 56950 & 5.00 & 21 \\
J1929+1154 & 56961 & 5.00 & 1 \\
J1929+1154 & 56963 & 5.00 & 1 \\
J1929+1154 & 56984 & 5.00 & 6 \\
J1929+1154 & 57017 & 5.00 & 3 \\
J1929+1154 & 57049 & 5.00 & 1 \\
\hline
J1929+1154 & Total & 284.71 & 298 \\
\hline
J2010+3147 & 54760$^{\ddagger}$ & 4.47 & 1 \\
J2010+3147 & 56375 & 19.00 & 48 \\
J2010+3147 & 56404 & 15.00 & 53 \\
J2010+3147 & 56406 & 15.00 & 31 \\
J2010+3147 & 56408 & 12.38 & 27 \\
J2010+3147 & 56410 & 6.60 & 10 \\
J2010+3147 & 56413 & 21.06 & 40 \\
J2010+3147 & 56418 & 15.07 & 31 \\
J2010+3147 & 56424 & 13.92 & 30 \\
J2010+3147 & 56436 & 15.00 & 44 \\
J2010+3147 & 56439 & 7.84 & 33 \\
J2010+3147 & 56464 & 11.83 & 43 \\
J2010+3147 & 56495 & 14.63 & 62 \\
J2010+3147 & 56518 & 9.00 & 35 \\
J2010+3147 & 56524 & 14.00 & 49 \\
J2010+3147 & 56584 & 7.87 & 15 \\
J2010+3147 & 56615 & 13.32 & 58 \\
J2010+3147 & 56641 & 10.03 & 50 \\
J2010+3147 & 56742 & 7.72 & 9 \\
J2010+3147 & 56771 & 0.04 & 0 \\
J2010+3147 & 56799 & 10.00 & 36 \\
J2010+3147 & 56800 & 10.00 & 12 \\
J2010+3147 & 56801 & 10.00 & 27 \\
J2010+3147 & 56803 & 10.00 & 16 \\
J2010+3147 & 56805 & 3.68 & 8 \\
J2010+3147 & 56808 & 10.00 & 16 \\
J2010+3147 & 56813 & 4.47 & 0 \\
J2010+3147 & 56818 & 10.00 & 29 \\
J2010+3147 & 56828 & 10.00 & 20 \\
J2010+3147 & 56885 & 10.00 & 29 \\
J2010+3147 & 56947 & 2.70 & 3 \\
\hline
J2010+3147 & Total & 324.65 & 865 \\
\hline
\end{longtable*}

\begin{longtable*}{c|c|c|c}
\caption{Details of each observation from the follow-up observations with CHIME.} \label{tab:chime-observation-details}\\\hline
Source & Date & Length & Number of pulses \\
& (MJD) & (minutes) & \\\hline\hline
\endhead
J0623+15 & 59216 & 14.39 & 0 \\
J0623+15 & 59219 & 14.39 & 0 \\
J0623+15 & 59221 & 14.39 & 0 \\
J0623+15 & 59222 & 14.43 & 0 \\
J0623+15 & 59223 & 14.39 & 0 \\
J0623+15 & 59224 & 14.39 & 0 \\
J0623+15 & 59226 & 14.43 & 0 \\
J0623+15 & 59227 & 14.27 & 0 \\
J0623+15 & 59228 & 14.43 & 0 \\
J0623+15 & 59229 & 13.43 & 0 \\
J0623+15 & 59231 & 13.47 & 0 \\
J0623+15 & 59235 & 14.39 & 0 \\
J0623+15 & 59236 & 14.43 & 0 \\
J0623+15 & 59241 & 14.38 & 0 \\
J0623+15 & 59242 & 14.47 & 0 \\
J0623+15 & 59243 & 14.39 & 0 \\
J0623+15 & 59246 & 13.39 & 0 \\
J0623+15 & 59247 & 14.39 & 0 \\
J0623+15 & 59249 & 14.43 & 0 \\
J0623+15 & 59251 & 14.34 & 0 \\
J0623+15 & 59254 & 14.39 & 0 \\
J0623+15 & 59255 & 13.43 & 0 \\
J0623+15 & 59257 & 14.47 & 0 \\
J0623+15 & 59259 & 14.39 & 0 \\
J0623+15 & 59262 & 14.43 & 0 \\
J0623+15 & 59264 & 14.39 & 0 \\
J0623+15 & 59265 & 14.39 & 0 \\
J0623+15 & 59267 & 14.39 & 0 \\
J0623+15 & 59268 & 14.39 & 0 \\
J0623+15 & 59269 & 14.43 & 0 \\
J0623+15 & 59270 & 13.47 & 0 \\
J0623+15 & 59271 & 14.39 & 0 \\
J0623+15 & 59273 & 14.34 & 0 \\
J0623+15 & 59274 & 14.39 & 0 \\
J0623+15 & 59277 & 14.38 & 0 \\
\hline
J0623+15 & Total & 499.09 & 0 \\
\hline
J0625+12 & 59215 & 14.23 & 0 \\
J0625+12 & 59216 & 14.19 & 0 \\
J0625+12 & 59217 & 14.23 & 0 \\
J0625+12 & 59218 & 14.23 & 0 \\
J0625+12 & 59220 & 14.27 & 0 \\
J0625+12 & 59221 & 14.23 & 0 \\
J0625+12 & 59225 & 14.19 & 0 \\
J0625+12 & 59226 & 14.19 & 0 \\
J0625+12 & 59227 & 14.27 & 0 \\
J0625+12 & 59228 & 13.82 & 0 \\
J0625+12 & 59230 & 13.19 & 0 \\
J0625+12 & 59233 & 14.27 & 0 \\
J0625+12 & 59235 & 14.23 & 0 \\
J0625+12 & 59236 & 14.23 & 0 \\
J0625+12 & 59237 & 14.27 & 0 \\
J0625+12 & 59243 & 14.23 & 0 \\
J0625+12 & 59244 & 14.23 & 0 \\
J0625+12 & 59245 & 14.19 & 0 \\
J0625+12 & 59246 & 14.27 & 0 \\
J0625+12 & 59247 & 14.27 & 0 \\
J0625+12 & 59249 & 14.19 & 0 \\
J0625+12 & 59251 & 13.79 & 0 \\
J0625+12 & 59252 & 14.19 & 0 \\
J0625+12 & 59253 & 14.23 & 0 \\
J0625+12 & 59254 & 14.19 & 0 \\
J0625+12 & 59255 & 14.27 & 0 \\
J0625+12 & 59259 & 14.19 & 0 \\
J0625+12 & 59260 & 14.23 & 0 \\
J0625+12 & 59261 & 14.27 & 0 \\
J0625+12 & 59266 & 14.27 & 0 \\
J0625+12 & 59268 & 14.27 & 0 \\
J0625+12 & 59269 & 14.27 & 0 \\
J0625+12 & 59270 & 13.75 & 0 \\
J0625+12 & 59271 & 14.27 & 0 \\
J0625+12 & 59272 & 14.23 & 0 \\
J0625+12 & 59274 & 13.75 & 0 \\
J0625+12 & 59275 & 14.27 & 0 \\
J0625+12 & 59276 & 13.82 & 0 \\
J0625+12 & 59277 & 14.27 & 0 \\
\hline
J0625+12 & Total & 552.00 & 0 \\
\hline
J1843+05 & 60679 & 13.99 & 0 \\
J1843+05 & 60681 & 13.99 & 0 \\
J1843+05 & 60686 & 13.67 & 0 \\
J1843+05 & 60688 & 13.99 & 0 \\
J1843+05 & 60689 & 13.99 & 0 \\
J1843+05 & 60692 & 13.99 & 0 \\
J1843+05 & 60698 & 13.99 & 0 \\
J1843+05 & 60702 & 13.99 & 0 \\
J1843+05 & 60710 & 13.99 & 0 \\
J1843+05 & 60713 & 13.99 & 0 \\
J1843+05 & 60718 & 13.99 & 0 \\
J1843+05 & 60748 & 13.99 & 0 \\
J1843+05 & 60757 & 13.99 & 0 \\
J1843+05 & 60765 & 13.99 & 0 \\
J1843+05 & 60766 & 13.99 & 0 \\
J1843+05 & 60775 & 13.99 & 0 \\
J1843+05 & 60787 & 13.95 & 0 \\
J1843+05 & 60823 & 13.99 & 0 \\
J1843+05 & 60835 & 13.99 & 0 \\
J1843+05 & 60836 & 13.99 & 0 \\
J1843+05 & 60873 & 13.99 & 0 \\
J1843+05 & 60897 & 13.99 & 0 \\
\hline
J1843+05 & Total & 307.46 & 0 \\
\hline
J1924+10 & 59005 & 7.10 & 0 \\
J1924+10 & 59006 & 4.82 & 0 \\
J1924+10 & 59008 & 14.07 & 0 \\
J1924+10 & 59009 & 14.07 & 0 \\
J1924+10 & 59010 & 14.06 & 0 \\
J1924+10 & 59011 & 14.07 & 0 \\
J1924+10 & 59013 & 14.11 & 0 \\
J1924+10 & 59014 & 14.07 & 0 \\
J1924+10 & 59015 & 14.01 & 0 \\
J1924+10 & 59016 & 14.03 & 0 \\
J1924+10 & 59017 & 14.07 & 0 \\
J1924+10 & 59018 & 14.07 & 0 \\
J1924+10 & 59019 & 14.06 & 0 \\
J1924+10 & 59020 & 14.03 & 0 \\
J1924+10 & 59021 & 14.03 & 0 \\
J1924+10 & 59022 & 14.07 & 0 \\
J1924+10 & 59023 & 14.07 & 0 \\
J1924+10 & 59024 & 14.06 & 0 \\
J1924+10 & 59025 & 14.07 & 0 \\
J1924+10 & 59026 & 14.07 & 0 \\
J1924+10 & 59027 & 14.03 & 0 \\
J1924+10 & 59028 & 13.91 & 0 \\
J1924+10 & 59029 & 14.03 & 0 \\
J1924+10 & 59038 & 14.11 & 0 \\
J1924+10 & 59053 & 14.07 & 0 \\
J1924+10 & 59060 & 14.14 & 0 \\
J1924+10 & 59063 & 13.79 & 0 \\
J1924+10 & 59074 & 14.07 & 0 \\
J1924+10 & 59090 & 13.59 & 0 \\
J1924+10 & 59092 & 14.07 & 0 \\
J1924+10 & 59107 & 13.59 & 0 \\
J1924+10 & 59109 & 14.07 & 0 \\
J1924+10 & 59111 & 14.07 & 0 \\
J1924+10 & 59115 & 14.06 & 0 \\
J1924+10 & 59119 & 14.11 & 0 \\
J1924+10 & 59121 & 14.11 & 0 \\
J1924+10 & 59138 & 14.11 & 0 \\
J1924+10 & 59144 & 13.19 & 0 \\
J1924+10 & 59148 & 14.11 & 0 \\
J1924+10 & 59164 & 13.15 & 0 \\
J1924+10 & 59165 & 13.22 & 0 \\
J1924+10 & 59191 & 14.07 & 0 \\
J1924+10 & 59199 & 13.87 & 0 \\
J1924+10 & 59207 & 13.23 & 0 \\
J1924+10 & 59218 & 13.59 & 0 \\
J1924+10 & 59228 & 13.87 & 0 \\
J1924+10 & 59231 & 13.59 & 0 \\
J1924+10 & 59240 & 13.19 & 0 \\
J1924+10 & 59249 & 14.11 & 0 \\
J1924+10 & 59259 & 13.87 & 0 \\
J1924+10 & 59265 & 13.51 & 0 \\
J1924+10 & 59270 & 13.19 & 0 \\
J1924+10 & 59275 & 14.07 & 0 \\
\hline
J1924+10 & Total & 720.81 & 0 \\
\end{longtable*}


\newpage

\end{document}